\documentclass[twocolumn,amsthm]{autart}

\usepackage{amsmath,amssymb}
\usepackage{booktabs,array}
\usepackage{graphicx}
\usepackage{xcolor}
\usepackage[authoryear,round]{natbib}
\makeatletter\@ifundefined{c@part}{\newcounter{part}}{}\makeatother
\usepackage[colorlinks=true,linkcolor=blue,citecolor=blue,urlcolor=blue]{hyperref}

\theoremstyle{plain}
\newtheorem{theorem}{Theorem}
\newtheorem{proposition}{Proposition}
\newtheorem{lemma}{Lemma}
\newtheorem{definition}{Definition}
\newtheorem{assumption}{Assumption}
\newtheorem{remark}{Remark}

\newcommand{\p}{{\rm p}}
\renewcommand{\e}{{\rm e}}
\newcommand{\dd}{\mathrm{d}}
\newcommand{\bd}{\bar\delta}
\newcommand{\bz}{\bar z}
\newcommand{\bzeta}{\bar\zeta}
\newcommand{\tu}{\tilde u}
\newcommand{\tv}{\tilde v}

\protected\def\Krstic{Krsti\'c}

\allowdisplaybreaks
\allowdisplaybreaks
\makeatletter\let\ps@copyright\ps@empty\makeatother

\begin{document}

\begin{frontmatter}

\title{Control of Underactuated Surface Vessels---Part II:\\ Underway Docking}

\author[UCSD]{Miroslav \Krstic}\ead{krstic@ucsd.edu},
\author[NTNU]{Kristin Y. Pettersen}\ead{kristin.y.pettersen@ntnu.no}

\address[UCSD]{Department of Mechanical and Aerospace Engineering, University of California San Diego, La Jolla, CA 92093, USA}
\address[NTNU]{Department of Engineering Cybernetics, Norwegian University of Science and Technology (NTNU), Trondheim, Norway}

\begin{abstract}
In underway directional docking, a second-order nonholonomic problem, six
pairs of states of two vehicles have to be made to match by two inputs.
The key challenge is to design time-invariant feedback of the relative states and
the leader's measured motion only, which makes the follower's convergence in range
outpaced, with no sway actuation whatsoever, by the bearing and heading
convergence of the follower relative to the leader's path. We begin with
nonmodular backstepping, choosing speed and commanded heading jointly to
control range and bearing. Curvature compensation and normalization of
angular errors by different powers of range remove the range singularity
from the model. Our construction passes from unicycle kinematics to the
ships' zero-sideslip points and through the sway dynamics to the physical
hulls. The design extends to a leader with a different hull
and to a docking point astern or abeam of the leader, where we show that,
while the leader turns, a follower without sway actuation holds the point
only with a heading that differs from the leader's. 
With strict control Lyapunov construction, we achieve global
$\kappa$-exponential convergence, with a rate independent of the initial error.
\end{abstract}

\end{frontmatter}

%\begingroup\footnotesize\setlength{\parskip}{0pt}
%\tableofcontents
%\endgroup
\setcounter{secnumdepth}{2}

\section{Introduction}

In the companion paper, \citet{pos} solved the problem of
stabilization/parking for underactuated surface vessels by static state
feedback in polar coordinates. In this paper we turn our attention to a
problem that differs fundamentally: parking to a target that is never
still, namely following another moving ship and docking to it, with
prescribed bearing, canonically taken to be from astern (directly behind).

\vspace*{-15pt}\paragraph{Underway docking.}
In underway docking \citep{bl}, a vessel joins another vessel that keeps
moving, as in the recovery of an unmanned surface vehicle. The follower
has to reach a docking point fixed to the leader and then move with it. We therefore require arrival from
astern, along the path of the docking point: the follower then reaches the point
at its velocity, and docking passes into
following without a further maneuver. The
follower has a surge force and a yaw moment as inputs,
no input in sway \citep{fossen}, and it slips sideways when it turns.

The same problem arises with autonomous underwater vehicles (AUVs).
An AUV recovered by a surface vessel under way \citep{sarda17} is usually recovered on the
surface, and two AUVs that dock under water generally do so at a common, fixed depth.
In both cases the docking takes place in a horizontal plane, with the
surge, sway and yaw equations of the
surface vessel. Our results therefore cover AUV--ASV,
AUV--AUV and ASV--ASV docking, under the conditions
of Remark~\ref{rem:auv}.

Underway docking is tracking of a vehicle whose speed stays above a
positive bound. It differs from stabilization at a fixed
pose, which for a vehicle is parking, and is not an extension of it.

In parking, the target is at rest and the vehicle has to stop. A vehicle
without a sideways control input, whether a cart that cannot slip or a vessel that slips but has no sway actuation, cannot be stopped at a pose by continuous
time-invariant feedback in Cartesian coordinates \citep{brockett,wichlund95}, and
this obstruction is the difficulty of parking. Time-varying
feedback stabilizes the vehicle at the pose \citep{mazenc,pettersen04,dongguo}, with an oscillatory approach. Polar coordinates remove
the obstruction \citep{pos}: the vehicle slows in proportion to its distance from the target,
so an error in its bearing or heading moves it less and less as it
approaches, and the angles need only converge.

In tracking a moving leader, the follower cannot slow down. There is
no obstruction: the motion along the leader's path makes the errors
controllable, and feedback in Cartesian error coordinates stabilizes them.
The difficulty is a different one. A follower that moves at the leader's speed $w$ with a
heading error $\theta$ drifts sideways at the rate $w\sin\theta$, however
close it is to the leader. For the follower to arrive from astern, and
not across the leader's path, this drift must not exceed the lateral
offset that remains, which is the range times the bearing. It suffices
that the bearing decay in proportion to the range, and the heading error
relative to the leader's path in proportion to its square.
Parking calls for no such rates; they are the reason for our
range-normalized states.

Neither problem contains the other. A parking control law brings the speed to
zero at the target, so it cannot follow a leader that keeps moving. A tracking control law keeps the leader's speed and uses the curvature $\kappa_\e=r_\e/w$ of the leader's path, which is unbounded as that speed goes to zero, so it cannot park.

One control law for both problems is possible \citep{djp}, and the goal
is natural: the vessel that docks and the vessel that holds a
pose are the same vessel with the same inputs. Such a control law,
however, carries the demands of each problem into the other. That of \citet{djp} parks in
Cartesian coordinates by a time-varying term, which remains
when the leader moves, where tracking does not require it. Behind a
slow leader the follower then zigzags where a smooth
approach from astern is available (Figure~\ref{fig:cmp}). This suggests that the two
problems are solved better by two dedicated control laws.

% \vspace*{-15pt}\paragraph{Other related literature.}
% Tracking of a reference ship by this underactuated vessel has been
% designed in Cartesian error coordinates. The control laws of \citet{pn,jiang}
% require a persistently exciting reference yaw rate, which excludes a
% straight course. This requirement is removed in \citet{djptac}, and
% \citet{djp} tracks a reference ship and stabilizes a fixed pose with one
% time-varying control law. The global control laws for stabilization at a fixed pose in
% Cartesian coordinates \citep{mazenc,pettersen04,dongguo} are periodic in
% time.

% In polar coordinates the unicycle is stabilized at a fixed pose by
% time-invariant feedback with strict Lyapunov functions
% \citep{part1,part2}. The bicycle \citep{bike} and the tractor-trailer
% \citep{tt} are stabilized exponentially in polar angles divided by the
% range. The vessel with sway dynamics is stabilized at a fixed pose in
% \citet{pos}, by steering its zero-sideslip point, which moves as a
% unicycle. For a moving unicycle whose turning is not known, \citet{dpe}
% gives pursuit with arrival along the evader's heading from favorable
% initial bearings, and \citet{wk} gives it from all initial bearings. This
% paper treats a leader whose motion is measured and a follower that slips
% sideways.

\vspace*{-15pt}\paragraph{What underway docking requires.}
Underway docking asks for five requirements at once:
(R1) the separation between the follower and a docking point fixed to the
leader converges to zero, not to a positive offset; the
docking point can be the leader's pivot point or a point astern or abeam of it,
such as a stern ramp or a crane recovery station;
(R2) the follower arrives from a prescribed direction, here from astern along
the path of the docking point;
(R3) it arrives with the velocity of the docking point;
(R4) these hold for a vessel with surge force and yaw moment as its only
inputs, whose sway is driven by its yaw rate, behind a leader that
turns freely and whose hull may differ from the follower's; and
(R5) the follower stays at the docking point while the leader turns.
The sway equation of a vessel without sway actuation is a second-order nonholonomic constraint, a
non-integrable relation between its sway acceleration, its sway and surge
speeds and its yaw rate \citep{oriolo91,wichlund95}, and it limits (R5):
a follower that holds a point fixed to a
turning leader with the leader's heading needs a sway motion that its
sway equation produces, for equal hulls, if and only if
\eqref{eq:off-cond} holds (Section~\ref{sec:vl}). Each
requirement has been achieved in the literature, but not all together.

\vspace*{-15pt}\paragraph{Tracking a reference ship: (R1) and (R3), (R4) in part, but not (R2) or (R5).}
In tracking control of underactuated ships the follower's position and heading
converge to those of a reference ship, so the separation and the relative
velocity vanish. Exponential tracking with experiments was obtained by
\citet{pn}, global tracking by Lyapunov's direct method
by \citet{jiang}, and global $\kappa$-exponential tracking by
cascaded designs \citep{lefeber03,leejiang04}. These results
require a reference yaw rate bounded away from zero
\citep{pn} or persistently exciting
\citep{jiang,lefeber03,leejiang04}, which excludes
a straight course; \citet{djptac} admit a straight-line reference if
the reference yaw rate decays while the surge speed stays away from
zero. 
Leader--follower formation tracking
\citep{wangloria24} has the same structure: each vessel tracks its predecessor
in a chain at a constant displacement in the earth-fixed frame, and only
the virtual vessel at its root, which generates the
reference, moves freely, with a persistently exciting yaw
rate. The unicycle trackers of \citet{jn} and \citet{panteley98} share this
structure, the latter requiring that the reference not converge to a
point or a straight line. In all of these results the objective
is convergence of the position and heading errors in Cartesian coordinates. The
direction from which the follower reaches the reference is not part of the
objective and is set by the initial condition and the gains
(Section~\ref{sec:jn} and Figure~\ref{fig:jn}, for the control law of \citet{jn}).
The reference is the motion of a ship, or of a point displaced from it in the
earth-fixed frame, so a docking point fixed to a turning leader, as in (R5),
does not arise.

\vspace*{-15pt}\paragraph{Formation at a prescribed separation: (R2) and (R3), but not (R1).}
Separation--bearing control \citep{desai01} regulates the distance
and the relative angle of a follower to a moving leader, and the bearing
kinematics contain the separation in the denominator.
\citet{consolini08} let the follower's relative position vary on
circular arcs in the leader's frame to respect input constraints. For ships, leader--follower
synchronization has been demonstrated in model-scale underway-replenishment
experiments \citep{kyrkjebo07}, and
underactuated surface vehicles have followed a virtual target at a desired
range from a master vessel in harbor experiments \citep{bibuli12}.
In all of these the desired separation is a positive constant. Docking is
excluded by the formulation, and the bearing dynamics become singular as the
separation tends to zero.

\vspace*{-15pt}\paragraph{Docking by a moving virtual target: (R1)--(R3) for a leader on a straight line.}
The natural way to dock with a formation or tracking controller is to move or
shrink the offset. In the underway docking procedure of \citet{bl}, the leader,
a mother ship, moves on a straight line, and the follower, an unmanned surface
vehicle, tracks a virtual target point by constant-bearing guidance. 
The target is placed on a
safety circle around the leader, moved along the circle to the
bearing of the docking point, and then moved inward to it, with
switching between the phases governed by synchronization variables.
The stability analysis assumes that the vehicle's velocity converges to the velocity
commanded by the guidance law, and the position error is then uniformly globally
asymptotically and locally exponentially stable. The procedure is demonstrated
in simulation with an underactuated vehicle, whose low-level surge speed and course controllers track the commanded velocity; the heading of the hull is not controlled. In the survey by
\citet{lexau23} of automated docking of marine surface
vessels from 1980 to 2023, this is the only paper on docking to a vessel
under way. \citet{kwon14} shrink a leader--follower docking
formation of kinematic unicycles on a prescribed schedule until the
docking surfaces meet, in simulation, without a
stability analysis. In these designs the direction of arrival is produced by
the motion of a virtual target, and the guarantee, where one is given, concerns
the position and velocity of the follower, not the heading and sideslip
of an underactuated hull at arrival, nor a leader that turns.

\vspace*{-15pt}\paragraph{Holding a point on a turning leader: (R5).}
In all of the designs above that dock at, or hold, a point away from the
leader's pivot point, either the leader moves on a straight line or the
follower can produce a sway force. The leader of
\citet{bl} and the moving
station in the simulation of \citet{kwon14} move on a straight line. In underway
replenishment the supply ship holds its station abeam of a turning main ship
with thrusters in sway; in the experiments of \citet{kyrkjebo07}
these thrusters saturate when the supply ship sails the outer curve. In
separation--bearing control of unicycles the follower's heading is not
prescribed \citep{desai01}, and \citet{consolini08}
let the follower's position vary on arcs in the leader's frame. Whether a vessel without sway actuation can hold a point
fixed to a turning leader with the leader's heading has not, to our knowledge,
been asked.

\vspace*{-15pt}\paragraph{Docking with an approach-angle constraint: (R1) and (R2), but not (R3) under way.}
\citet{verma25} dock an underactuated surface vessel at a
stationary dock with a prescribed approach angle, by sliding-mode control of
the range and the line-of-sight angle with full vessel dynamics; the
line-of-sight dynamics contain the range in the denominator, and the target
speed at docking is zero.
\citet{yadegar25} dock a kinematic nonholonomic robot to a target
that moves at constant velocity, with prescribed orientation and closing
velocity, by a modified proportional navigation law whose gains are computed
from the initial engagement so that terminal conditions hold at a finite
interception time. None of these designs treats a leader that turns, or gives a
Lyapunov certificate for the closed loop of an underactuated vessel behind a
moving leader.

\vspace*{-15pt}\paragraph{Guidance with a terminal heading constraint: (R2), and (R3) without guarantees.}
Arrival from a prescribed direction, together with matching of the target's
velocity, is the objective of rendezvous guidance. For aerial recovery of UAVs
with a cable-drogue, \citet{thakar16} use planar kinematics of unicycle type,
with the lateral acceleration as turning input, and combine a terminal heading
constraint with speed matching; for a drogue that moves on a straight line they
compare pure-pursuit and trajectory-shaping guidance in simulation by the
initial conditions that lead to rendezvous. For the unicycle, \citet{dpe} and
\citet{wk} prescribe the arrival direction by feedback with strict Lyapunov
functions, but the pursuer reaches the evader with mismatched speed.

% The gap statement and its mapping to the contributions
\vspace*{-15pt}\paragraph{Gap.}
No existing design achieves (R1)--(R5) together. Tracking controllers reach
zero separation with matched velocity but leave the direction of arrival free.
Formation controllers prescribe the direction but keep a positive separation.
Virtual-target procedures reach the leader from a prescribed direction with
matched velocity, but for a leader on a straight line and with an analysis
that does not cover the heading of the hull. Approach-angle docking and
terminal-heading guidance prescribe the direction at zero separation, but either
the target is at rest, or the target moves at constant velocity and the vehicle
is kinematic, or the arrival is an interception at nonzero relative velocity. None of them addresses a docking
point away from the pivot point of a turning leader for a follower without sway
actuation.

\vspace*{-15pt}\paragraph{Contributions.}
This paper closes this gap as follows.
\begin{itemize}

\item {\em Range singularity removed from docking model.}
The bearing, which encodes the direction of arrival, is defined through a division by the range. Polar parking \citep{pos,part1} removes the division by making speed vanish with range. Under way the follower's speed stays near the leader's, and this mechanism is unavailable. We resolve this difficulty by incorporating the leader's curvature and the relative decay rates of the angular errors into the state variables. Beyond the range-normalized parking coordinates of \citet{bike,tt}, which divide the polar angles by the range for a target at rest, our construction combines curvature compensation with different powers of range so that the transformed equations remain well defined at zero separation and nonzero forward speed. This exposes a strict Lyapunov design for the docking dynamics and allows us to carry the directional guarantee from unicycle kinematics to a ship whose only control inputs are surge force and yaw moment. The normalizing weight saturates with the range, and the range enters the strict Lyapunov function through the integral of that weight, so that the closing speed is bounded at any range.

\item {\em Exponential docking with prescribed direction of arrival and the leader's velocity.}
The Cartesian tracking guarantees of \citet{pn,jiang,djptac,djp} establish convergence to a reference motion but do not specify the direction from which the follower reaches it. Directional unicycle pursuit \citep{dpe,wk} prescribes the direction but arrives with mismatched speed. Here, the direction of arrival is part of the control objective together with the leader's velocity. Nonmodular backstepping in the normalized states gives time-invariant feedback of the relative states and the measured leader motion, with a strict Lyapunov
function, under which the hulls dock while their zero-sideslip points arrive from astern, and the follower's speed and turning rate converge to the leader's (Theorems~\ref{thm:uni}--\ref{thm:dyn}, Proposition~\ref{prop:par}). The convergence is global $\kappa$-exponential in the sense of \citet{sordalen95}, with a decay rate set by the gains and independent of the initial error. The design requires neither persistent turning of the leader nor an imposed periodic excitation. Directional arrival is thereby made compatible with settling onto the leader's continuing motion, through bounded surge-force and yaw-moment inputs.

\item \emph{From unicycle kinematics to the underactuated hull.} Through the
zero-sideslip points and the sway dynamics, the guarantee is carried from
unicycles (Theorem~\ref{thm:uni}) to the vessel with surge speed and yaw rate as
inputs (Theorem~\ref{thm:kin}), to the vessel with surge force and yaw moment as
inputs (Theorem~\ref{thm:dyn}), and to a leader with a different hull
(Theorem~\ref{thm:vl}), which covers an AUV that docks with a surface vessel
(Remark~\ref{rem:auv}).

\item \emph{Docking at an offset point, and what a follower without sway
actuation cannot do.} The follower docks at any point fixed to the leader, at a
distance $d$ and an angle $\beta$ from its heading, astern or abeam, behind a
leader with a different hull (Theorem~\ref{thm:vl}). When the leader turns,
such a follower cannot hold both the offset point and the leader's heading
unless \eqref{eq:off-cond} holds, which it does without offset and on a
straight course. It holds the point, with a heading that differs from the
leader's by an angle $\eta$ bounded in terms of the leader's turning. This is
the limit that the second-order nonholonomic sway constraint places on underway
docking.

\end{itemize}
Theorems~\ref{thm:uni}--\ref{thm:dyn} give the control law for unicycles,
for surge speed and yaw rate inputs, and for surge force and yaw moment
inputs. Proposition~\ref{prop:par} gives the
follower's heading at arrival, and Theorem~\ref{thm:vl} treats docking at an offset point and the leader
with a different hull. Throughout, the leader's speed stays above a positive bound; a target at rest is treated in \citet{pos}.

\subsection*{Nomenclature}
\addcontentsline{toc}{subsection}{Nomenclature}

Section~\ref{sec:uni} uses $\rho$, $\delta$, $\gamma$, $w_\p$, $\omega_\p$,
$w_\e$, $\omega_\e$ for two unicycles. For the zero-sideslip points of two
ships these are $\rho_a$, $\delta_a$, $\gamma_a$, $u_a$, $r_\p$, $w$,
$r_\e$.

\begin{list}{}{\setlength{\labelwidth}{0.24\columnwidth}\setlength{\labelsep}{0.02\columnwidth}\setlength{\leftmargin}{0.26\columnwidth}\setlength{\itemsep}{1pt}\setlength{\parsep}{0pt}\setlength{\topsep}{3pt}\raggedright}
\item[${\rm p},\ {\rm e},\ a$\hfill] Follower; leader; zero-sideslip point.
\vspace{3pt}\item[$\rho$\hfill] Distance between the pivot points of the two ships.
\item[$\delta$\hfill] Bearing: angle from the leader's heading to the line of sight.
\item[$\gamma$\hfill] Line-of-sight angle: angle from the follower's heading to the line of sight.
\item[$\theta$\hfill] Heading error, $\theta=\delta-\gamma{}=\psi_\p-\psi_\e$, with $\psi_\p$ and $\psi_\e$ the headings of the follower and the leader.
\item[$u_\p,\ v_\p,\ r_\p$\hfill] Surge speed, sway speed and yaw rate of the follower.
\item[$u_\e,\ v_\e,\ r_\e$\hfill] Surge speed, sway speed and yaw rate of the leader.
\item[$u_m,\ u_M,\ a_M$\hfill] Lower and upper bounds on $u_\e$; bound on the leader's surge acceleration.
\item[$\tau_u,\ \tau_r$\hfill] Mass-normalized surge force and yaw moment of the follower.
\item[$X(u),\ Y_0$\hfill] Sway acceleration per unit yaw rate, $X(u)=X_0+X_1u$, $X_1<0$; sway damping rate, $Y_0<0$.
\item[$\mu,\ \omega_M$\hfill] $\mu=\min\{Y_0/X_1,-Y_0\}$; bound on the leader's yaw rate.
\vspace{3pt}\item[$\rho_a,\ \delta_a,\ \gamma_a$\hfill] Distance, bearing and line-of-sight angle of the zero-sideslip points.
\item[$u_a,\ w$\hfill] Forward speeds of the follower's and of the leader's zero-sideslip point.
\item[$u_{\min},\ u_{\max}$\hfill] Lower and upper bounds on $w$.
\item[$\kappa_\e$\hfill] Curvature of the leader's path, $\kappa_\e=r_\e/w$.
\vspace{3pt}\item[$\ell,\ m,\ \sigma$\hfill] Normalization length; weight $m=\ell\rho_a/(\rho_a+\ell)$; $\sigma=m/\ell$.
\item[$\bar\delta,\ \bar\theta$\hfill] Normalized bearing $\delta_a/m$; normalized heading error $(\theta+\rho_a\kappa_\e)/m^2$.
\item[$\bar u,\ \bar r$\hfill] Normalized closing rate $(u_a-w)/m$; normalized yaw rate, the rate of the heading error divided by $m^2$.
\item[$\bar\alpha$\hfill] Normalized yaw acceleration.
\item[$a,\ b$\hfill] Components of the commanded velocity along and across the line of sight.
\item[$z,\ r^\ast$\hfill] Angle from the commanded velocity to the follower's heading; yaw-rate reference.
\item[$k_1,\ k_2,\ c,\ k_r$\hfill] Positive gains.
\item[$M,\ \nu$\hfill] Normalized docking error; its decay rate.
\end{list}

\section{Unicycle following}\label{sec:uni}

Two unicycles move in the plane: a follower with speed $w_\p$ and steering
rate $\omega_\p$, the control inputs, and a leader with speed
$w_\e(t)$ and steering rate $\omega_\e(t)$. With the range $\rho$, the
bearing $\delta$, the line-of-sight angle $\gamma$ and the heading error
$\theta=\delta-\gamma$ of the Nomenclature, for $\rho>0$, as in \citet{dpe},
\begin{subequations}\label{eq:uni}
\begin{align}
\dot\rho&=w_\e(t)\cos\delta-w_\p\cos\gamma,\label{eq:uni-rho}\\
\dot\delta&=\frac{w_\p\sin\gamma-w_\e(t)\sin\delta}{\rho}-\omega_\e(t),\label{eq:uni-delta}\\
\dot\gamma&=\frac{w_\p\sin\gamma-w_\e(t)\sin\delta}{\rho}-\omega_\p .\label{eq:uni-gamma}
\end{align}
\end{subequations}

\subsection{Range-normalized model}\label{sec:norm}

The curvature of the leader's path is
\begin{equation}
\kappa_\e(t)=\frac{\omega_\e(t)}{w_\e(t)} .
\label{eq:kappa}
\end{equation}
With a length $\ell>0$, let
\begin{subequations}\label{eq:nvar}
\begin{align}
\bar\delta&=\frac{\delta}{m},\label{eq:nd}\\
\bar\theta&=\frac{\theta+\rho\,\kappa_\e(t)}{m^2},\label{eq:nt}\\
\bar u&=\frac{w_\p-w_\e(t)}{m},\label{eq:nu}\\
\bar r&=\frac{1}{m^2}\Bigl(\omega_\p-\omega_\e(t)+\frac{\dd}{\dd t}\bigl(\rho\,\kappa_\e(t)\bigr)\Bigr),\label{eq:nr}\\
m&=\frac{\ell\,\rho}{\rho+\ell},\label{eq:m}\\
\sigma&=\frac{\rho}{\rho+\ell}.\label{eq:sig}
\end{align}
\end{subequations}
In the normalized states $(\rho,\bar\delta,\bar\theta)$ and
inputs $(\bar u,\bar r)$, \eqref{eq:uni} is
\begin{subequations}\label{eq:nmodel}
\begin{align}
\dot\rho&=-m\,\bar u\cos\gamma+m^2w_\e(t)\,e_1,\label{eq:n1}\\
\dot{\bar\delta}&=(1-\sigma)\bigl[-w_\e(t)\,\bar\theta+\bar u\,\bigl(\bar s+(1-\sigma)\,\bar\delta\cos\gamma\bigr)\bigr]\notag\\
&\quad+(1-\sigma)\,m\,w_\e(t)\bigl(e_2-(1-\sigma)\,\bar\delta\,e_1\bigr),\label{eq:n2}\\
\dot{\bar\theta}&=\bar r+2(1-\sigma)^2\bigl(\bar u\cos\gamma-m\,w_\e(t)\,e_1\bigr)\bar\theta ,\label{eq:n3}
\end{align}
\end{subequations}
where
\begin{equation}
\begin{gathered}
\bar s=\frac{\sin\gamma}{m},\qquad
e_1=\frac{\cos\delta-\cos\gamma}{m^2},\\
e_2=\frac{\sin\gamma-\sin\delta+\theta}{m^3}
\end{gathered}
\label{eq:naux}
\end{equation}
are smooth in $(\rho,\bar\delta,\bar\theta)$ for $\rho\ge0$.
For $\rho\ll\ell$, \eqref{eq:nmodel} is
\begin{subequations}\label{eq:near}
\begin{align}
\dot\rho&=-\bar u\,\rho,\label{eq:near1}\\
\dot{\bar\delta}&=-w_\e(t)\,\bar\theta+\bar u\,\bigl(2\bar\delta+\kappa_\e(t)\bigr),\label{eq:near2}\\
\dot{\bar\theta}&=\bar r+2\bar u\,\bar\theta .\label{eq:near3}
\end{align}
\end{subequations}

The state $\bar\delta$ is a curvature: the circle through the follower's
position that reaches the leader's position tangent to the leader's
heading has curvature $-2\sin\delta/\rho$, which is $-2\bar\delta$ for
$\rho\ll\ell$. To first order in $\rho$, the angle $-\rho\,\kappa_\e$ is
the heading of the leader's path at the distance $\rho$ behind the leader,
so $\bar\theta$ is the follower's heading error relative to the leader's
path, per square of the range. The input $\bar u$ is the rate at which the
range closes.

The right sides of \eqref{eq:nmodel} and \eqref{eq:near} contain no
division by $\rho$, whereas \eqref{eq:uni-delta} and \eqref{eq:uni-gamma}
do: the division is carried by the states $\bar\delta$ and $\bar\theta$. The terms
$2\bar u\,\bar\delta$ and $2\bar u\,\bar\theta$ in \eqref{eq:near} state
that a fixed lateral offset and a fixed heading error are an arrival arc of
growing curvature as the range closes.

The weight $m$ saturates. Near the leader it is the range and removes
the division by $\rho$. Far from the leader it is the constant $\ell$ and
bounds the commands: by \eqref{eq:n1}, $\bar u$ closes the range at the
speed $m\,\bar u\cos\gamma$, at most $\ell\,|\bar u|$, and by
\eqref{eq:nr} the steering rate relative to the leader's path
is $m^2\,\bar r$, at most $\ell^2\,|\bar r|$. With $\rho$ in
place of $m$, bounded inputs $\bar u$ and $\bar r$ would be a
speed proportional to the range and a steering rate proportional to the
range squared.

\subsection{Control law and stability estimate (Theorem \ref{thm:uni})}

Next we employ the {\em nonmodular} backstepping approach
of \citet{part2}: the control $\bar u$ in \eqref{eq:n1} and the virtual
control $\bar\theta$ in \eqref{eq:n2} are chosen jointly, as
$\bar u=(\sqrt{a^2+b^2}-w_\e)/m$ and
$\bar\theta=(\delta-\arctan(b/a)+\rho\,\kappa_\e)/m^2$,
where $a$ and $b$ are the components of a commanded velocity along and
across the line of sight.

\begin{theorem}[Exponential unicycle docking from astern]\label{thm:uni}
Let $w_\e$ and $\omega_\e$ be continuous, with bounded, piecewise
continuous derivatives, and let $0<w_{\min}\le w_\e(t)\le w_{\max}$ and
$|\omega_\e(t)|\le\omega_M$ for all $t\ge0$. Let the follower measure
$\rho$, $\delta$, $\gamma$, $w_\e$, $\omega_\e$, $\dot w_\e$ and
$\dot\omega_\e$. Consider \eqref{eq:uni} under the feedback
\begin{subequations}\label{eq:ulaw}
\begin{align}
w_\p&=\sqrt{a^2+b^2},\label{eq:uw}\\
\omega_\p&=\omega_\e(t)+\dot\delta-\frac{\dd}{\dd t}\arctan\frac ba\notag\\
&\quad+\Bigl(c-\frac{2(1-\sigma)\dot\rho}{\rho}\Bigr)z+m^2\chi,\label{eq:uom}
\end{align}
\end{subequations}
where
\begin{subequations}\label{eq:where}
\begin{align}
z&=\gamma-\arctan\frac ba,\label{eq:z}\\
a&=w_\e(t)+k_1m,\label{eq:a}\\
b&=w_\e(t)\sin\delta+\rho\,\omega_\e(t)-k_2\,m\,\delta\notag\\
&\quad-(1-\sigma)\,\delta\,\bigl[k_1m+w_\e(t)(1-\cos\delta)\bigr],\label{eq:b}\\
\chi&=\sigma\,m^2E_1+\frac{\delta}{\rho}\bigl(E_2-(1-\sigma)\,\delta\,E_1\bigr),\label{eq:chi}\\
E_1&=\frac{a(1-\cos z)+b\sin z}{z},\notag\\
E_2&=\frac{a\sin z-b(1-\cos z)}{z},\label{eq:E}
\end{align}
\end{subequations}
$m$ and $\sigma$ are given by \eqref{eq:m}, \eqref{eq:sig}, the derivatives in
\eqref{eq:uom} are taken along \eqref{eq:uni-rho}, \eqref{eq:uni-delta}
with \eqref{eq:uw}, and the gains $k_1$, $k_2$, $c$ are positive. Let the
normalized tracking error be
\begin{equation}
\begin{gathered}
M=\rho+|\bar\delta|+|\bar\theta-\bar\theta^\ast(\rho,t)|,\\
\bar\theta^\ast(\rho,t)=\frac{k_1\,\kappa_\e(t)}{(1-\sigma)\bigl(w_\e(t)+k_1m\bigr)},
\end{gathered}
\label{eq:Mu}
\end{equation}
and let $0<\nu<\min\{k_1,k_2,c\}$. Then there are $\kappa,\kappa_\omega\in\mathcal K_\infty$ such that, for every initial
condition with $\rho(0)>0$ and all $t\ge0$, the closed-loop solution exists, is unique,
and satisfies $\rho(t)>0$ and
\begin{subequations}\label{eq:uclaims}
\begin{align}
M(t)&\le\kappa\bigl(M(0)\bigr)\,e^{-\nu t},\label{eq:uclaim}\\
|w_\p(t)-w_\e(t)|+|\omega_\p(t)-\omega_\e(t)|&\le\kappa_\omega\bigl(M(0)\bigr)\,e^{-\nu t}.
\label{eq:uclaim2}
\end{align}
\end{subequations}
\end{theorem}

The speed \eqref{eq:uw} as the norm of a commanded velocity is from
\citet{wk}. The proof, in Appendix~\ref{app:proof}, uses the
backstepping target system, with $\bar z=z/m^2$,
\begin{subequations}\label{eq:utarget}
\begin{align}
\dot\rho&=-k_1m-w_\e(t)\bigl(1-\cos(m\bar\delta)\bigr)+m^2\,\bar z\,E_1,\label{eq:ut1}\\
\dot{\bar\delta}&=-k_2(1-\sigma)\,\bar\delta+\bar z\,(1-\sigma)\bigl(E_2-(1-\sigma)\,m\,\bar\delta\,E_1\bigr),\label{eq:ut2}\\
\dot{\bar z}&=-c\,\bar z-\chi .\label{eq:ut3}
\end{align}
\end{subequations}
and the strict Lyapunov function
\begin{equation}
\begin{aligned}
V&=\Phi(\rho)+\tfrac12\bigl(\bar\delta^2+\bar z^2\bigr),\\
\Phi(\rho)&=\rho-\ell\ln\Bigl(1+\frac{\rho}{\ell}\Bigr),
\end{aligned}
\label{eq:uV}
\end{equation}
whose derivative along \eqref{eq:utarget} is
\begin{equation}
\begin{aligned}
\dot V&=-\sigma\bigl[k_1m+w_\e(t)(1-\cos\delta)\bigr]\\
&\quad-k_2(1-\sigma)\,\bar\delta^2-c\,\bar z^2 .
\end{aligned}
\label{eq:uVdot}
\end{equation}
The function $\Phi$, defined by $\Phi'=\sigma$, is quadratic for
$\rho\ll\ell$ and linear for $\rho\gg\ell$. With this choice, the
term of $\chi$ in \eqref{eq:chi} that cancels the cross term of the range
in $\dot V$ is $\sigma m^2E_1$, in which $\sigma m^2\le\ell^2$. With $\rho$ as the weight and a quadratic range term in $V$, this
term would be $\rho^3E_1$. With the saturation, by \eqref{eq:ut1}, the range closes far from the
leader at the constant speed $k_1\ell$, and the decay is exponential only
near the leader. The function $\kappa$ in \eqref{eq:uclaim} quantifies the
absence of a globally exponential decay with an initial condition-dependent
overshoot.

\subsection{Comparison with tracking in Cartesian coordinates}\label{sec:jn}

\citet{jn} design control laws with which a unicycle tracks a reference
unicycle, the leader of this section. Their objective is
convergence to zero of the position error in the follower's
frame, with the components $\rho\cos\gamma$ and $\rho\sin\gamma$, and of
the heading error $\theta$. Their global tracking control law
\citep[Proposition~3]{jn} is
\begin{subequations}\label{eq:jn}
\begin{align}
w_\p&=w_\e(t)\cos\theta+c_x\,\rho\cos\gamma,\label{eq:jnw}\\
\omega_\p&=\omega_\e(t)+w_\e(t)\,\rho\sin\gamma\,\frac{\sin\theta}{\theta}-c_\theta\,\theta,\label{eq:jnom}
\end{align}
\end{subequations}
with gains $c_x,c_\theta>0$. With $V=\tfrac12\bigl(\rho^2+\theta^2\bigr)$,
\eqref{eq:jn} gives
$\dot V=-c_x\,\rho^2\cos^2\gamma-c_\theta\,\theta^2$, and, for a leader as
in Theorem~\ref{thm:uni}, $\rho(t)\to0$ and $\theta(t)\to0$ for all initial
conditions.

Theorem~\ref{thm:uni} differs from this result in objective and
guarantee. The convergence of $\rho$ and $\theta$ to zero leaves free the
direction from which the follower reaches the leader. By \eqref{eq:uclaim}, the bearing per unit of
range, $\bar\delta$, converges to zero, which is arrival from astern.
Under \eqref{eq:jn}, $\dot V$ is negative semidefinite, and the
convergence follows by Barbalat's lemma, without a decay rate. The estimate
\eqref{eq:uclaim} is exponential, with the rate $\nu$ set by the gains.

\begin{figure}[t]
\centering
\includegraphics[width=0.85\columnwidth]{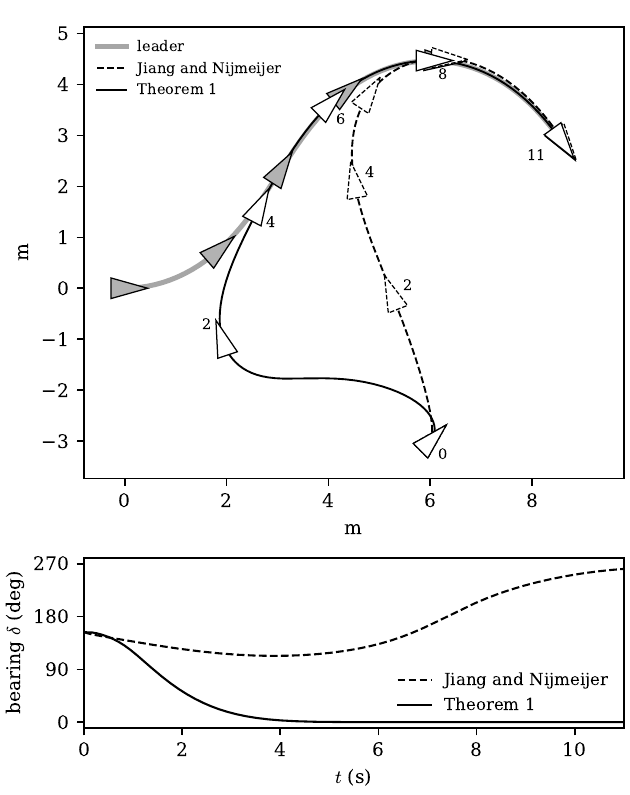}
\caption{Unicycle following under the control law \eqref{eq:jn} of
\citet{jn} (dashed; $c_x=0.5$, $c_\theta=2$) and of
Theorem~\ref{thm:uni} (solid; $k_1=0.5$, $k_2=1$, $c=2$, $\ell=2$~m), for
$w_\e=1$~m/s, $\omega_\e=0.4\cos(0.4t)$~rad/s. Top: paths, with the
vehicles at the marked times (s). Bottom: bearing $\delta$.}
\label{fig:jn}
\end{figure}

In Figure~\ref{fig:jn} the leader turns alternately left and right,
and the follower starts ahead of it. Both followers reach the leader. Under \eqref{eq:jn} the
bearing stays away from zero: the follower reaches the leader from the
side and crosses ahead of it. Under the control law of
Theorem~\ref{thm:uni} the bearing converges to zero early,
and the follower closes along the leader's path,
from astern.

\section{Ship model, zero-sideslip points and normalized variables}\label{sec:model}

The ship model is that of \citet{pos}. The leader's sway speed obeys
\begin{equation}
\dot v_\e=Y_0v_\e+X(u_\e(t))\,r_\e(t),\qquad X(u)=X_0+X_1u .
\label{eq:leader}
\end{equation}
With $\rho$, $\delta$, $\gamma$ the polar variables of
Section~\ref{sec:uni} for the pivot points of the two ships, for $\rho>0$,
\begingroup\medmuskip=1.5mu\thickmuskip=2.5mu\relax
\begin{subequations}\label{eq:orig}
\begin{align}
\dot\rho&=-u_\p\cos\gamma-v_\p\sin\gamma+u_\e(t)\cos\delta+v_\e(t)\sin\delta,\label{eq:o1}\\
\dot\delta&=\frac{u_\p\sin\gamma-v_\p\cos\gamma-u_\e(t)\sin\delta+v_\e(t)\cos\delta}{\rho}-r_\e(t),\label{eq:o2}\\
\dot\gamma&=\frac{u_\p\sin\gamma-v_\p\cos\gamma-u_\e(t)\sin\delta+v_\e(t)\cos\delta}{\rho}-r_\p,\label{eq:o3}\\
\dot u_\p&=f_u(u_\p,v_\p,r_\p)+\tau_u,\label{eq:o4}\\
\dot v_\p&=Y_0v_\p+X(u_\p)\,r_\p,\label{eq:o5}\\
\dot r_\p&=f_r(u_\p,v_\p,r_\p)+\tau_r.\label{eq:o6}
\end{align}
\end{subequations}
\endgroup
With the inputs $(u_\p,r_\p)$, surge speed and yaw rate, the
model is \eqref{eq:o1}--\eqref{eq:o3}, \eqref{eq:o5}. With the inputs
$(\tau_u,\tau_r)$, surge force and yaw moment, the model is
\eqref{eq:orig}.

\begin{assumption}\label{as:hull}
The two vessels have equal hulls: \eqref{eq:leader} and \eqref{eq:o5} hold
with the same $X_0$, $X_1<0$, $Y_0<0$.
\end{assumption}

\begin{definition}[Zero-sideslip point; \citealp{pos}]\label{def:zsp}
The \emph{zero-sideslip point} of a vessel is the point displaced from its
pivot point by $X(u)/Y_0$ along the forward axis and by $-v/Y_0$ along the
lateral axis, which is the forward axis rotated counterclockwise by
$90^\circ$.
\end{definition}

\begin{remark}[Underwater vehicles]\label{rem:auv}
\rm The model \eqref{eq:orig} also describes an AUV in the horizontal plane,
either on the surface or at a constant depth maintained by a separate depth
controller, with roll and pitch small \citep{fossen}. The surge force is
produced by the propeller and the yaw moment by the rudder. The sway force
that the rudder also produces does not appear in \eqref{eq:o5}, since the
model is written at the pivot point, as for the surface vessel
\citep{pos}. The
hulls of an AUV and a surface vessel differ, so Assumption~\ref{as:hull}
does not hold for AUV--ASV docking (Section~\ref{sec:vl}).
\end{remark}

The two zero-sideslip points have the polar variables
\begin{subequations}\label{eq:zspdef}
\begin{align}
\rho_a&=\sqrt{q_x^2+q_y^2},\qquad\gamma_a=\arg(q_x,q_y),\qquad\delta_a=\gamma_a+\theta,\label{eq:zsp}\\
q_x&=\rho\cos\gamma+\frac{X(u_\e(t))}{Y_0}\cos\theta-\frac{v_\e(t)}{Y_0}\sin\theta-\frac{X(u_\p)}{Y_0},\label{eq:qx}\\
q_y&=\rho\sin\gamma-\frac{X(u_\e(t))}{Y_0}\sin\theta-\frac{v_\e(t)}{Y_0}\cos\theta+\frac{v_\p}{Y_0}.\label{eq:qy}
\end{align}
\end{subequations}
The leader's zero-sideslip point moves along the leader's heading at the
speed
\begin{equation}
w(t)=u_\e(t)+\frac{v_\e(t)\,r_\e(t)}{Y_0}+\frac{X_1}{Y_0}\,\dot u_\e(t).
\label{eq:w}
\end{equation}
As in \citet{pos}, that of the follower moves along the
follower's heading, at the speed
$u_a=u_\p+v_\p r_\p/Y_0+(X_1/Y_0)\,\dot u_\p$, and for $\rho_a>0$,
\eqref{eq:orig} is
\begin{subequations}\label{eq:zmodel}
\begin{align}
\dot\rho_a&=w(t)\cos\delta_a-u_a\cos\gamma_a,\label{eq:z1}\\
\dot\delta_a&=\frac{u_a\sin\gamma_a-w(t)\sin\delta_a}{\rho_a}-r_\e(t),\label{eq:z2}\\
\dot\gamma_a&=\frac{u_a\sin\gamma_a-w(t)\sin\delta_a}{\rho_a}-r_\p,\label{eq:z3}\\
\dot u_\p&=\frac{Y_0}{X_1}\bigl(u_a-u_\p\bigr)-\frac{v_\p\,r_\p}{X_1},\label{eq:z4}\\
\dot v_\p&=Y_0v_\p+X(u_\p)\,r_\p,\label{eq:z5}\\
\dot r_\p&=f_r(u_\p,v_\p,r_\p)+\tau_r.\label{eq:z6}
\end{align}
\end{subequations}
The first three lines are the unicycle model \eqref{eq:uni} with
$(w_\e,\omega_\e,w_\p,\omega_\p)$ replaced by $(w,r_\e,u_a,r_\p)$. With
surge speed and yaw rate inputs, the inputs of
\eqref{eq:z1}--\eqref{eq:z5} are $(u_a,r_\p)$, and $u_\p$ is a state of the
controller. With surge force and yaw moment inputs, $u_a$ is assigned by
$\tau_u=-f_u+(Y_0/X_1)(u_a-u_\p)-v_\p r_\p/X_1$.

With the curvature $\kappa_\e=r_\e/w$ of the path of the leader's
zero-sideslip point and $\ell$ of the order of the hull length, the normalized variables for the zero-sideslip points, defined as in \eqref{eq:nvar}, are
\begin{subequations}\label{eq:znvar}
\begin{align}
\bar\delta&=\frac{\delta_a}{m},\label{eq:znd}\\
\bar\theta&=\frac{\theta+\rho_a\,\kappa_\e(t)}{m^2},\label{eq:znt}\\
\bar u&=\frac{u_a-w(t)}{m},\label{eq:znu}\\
\bar r&=\frac{1}{m^2}\Bigl(r_\p-r_\e(t)+\frac{\dd}{\dd t}\bigl(\rho_a\,\kappa_\e(t)\bigr)\Bigr),\label{eq:znr}\\
\bar\alpha&=\frac{1}{m^2}\Bigl(f_r(u_\p,v_\p,r_\p)+\tau_r-\dot r_\e(t)+\frac{\dd^2}{\dd t^2}\bigl(\rho_a\,\kappa_\e(t)\bigr)\Bigr),\label{eq:zna}\\
m&=\frac{\ell\,\rho_a}{\rho_a+\ell},\label{eq:zm}\\
\sigma&=\frac{\rho_a}{\rho_a+\ell},\label{eq:zsig}
\end{align}
\end{subequations}
and the model \eqref{eq:zmodel} is
\begin{subequations}\label{eq:znmodel}
\begin{align}
\dot\rho_a&=-m\,\bar u\cos\gamma_a+m^2w(t)\,e_1,\label{eq:zn1}\\
\dot{\bar\delta}&=(1-\sigma)\bigl[-w(t)\,\bar\theta+\bar u\,\bigl(\bar s+(1-\sigma)\,\bar\delta\cos\gamma_a\bigr)\bigr]\notag\\
&\quad+(1-\sigma)\,m\,w(t)\bigl(e_2-(1-\sigma)\,\bar\delta\,e_1\bigr),\label{eq:zn2}\\
\dot{\bar\theta}&=\bar r+2(1-\sigma)^2\bigl(\bar u\cos\gamma_a-m\,w(t)\,e_1\bigr)\bar\theta ,\label{eq:zn3}\\
\dot u_\p&=\frac{Y_0}{X_1}\bigl(w(t)+m\,\bar u-u_\p\bigr)-\frac{v_\p\,r_\p}{X_1},\label{eq:zn4}\\
\dot v_\p&=Y_0v_\p+X(u_\p)\,r_\p,\label{eq:zn5}\\
\dot{\bar r}&=\bar\alpha+2(1-\sigma)^2\bigl(\bar u\cos\gamma_a-m\,w(t)\,e_1\bigr)\bar r ,\label{eq:zn6}
\end{align}
\end{subequations}
with $\bar s$, $e_1$, $e_2$ given by \eqref{eq:naux} for $\delta_a$,
$\gamma_a$, and with
$r_\p=r_\e(t)-\frac{\dd}{\dd t}(\rho_a\kappa_\e(t))+m^2\,\bar r$ in \eqref{eq:zn4},
\eqref{eq:zn5}. The inputs are $(\bar u,\bar r)$ for the first five lines
and $(\bar u,\bar\alpha)$ for all six, where $\bar\alpha$ contains the
derivative of $\bar u$. For $\rho_a\ll\ell$, the first three lines are \eqref{eq:near}
with $(\rho,w_\e)$ replaced by $(\rho_a,w)$.

\begin{figure*}[t]
\centering
\includegraphics[width=0.85\textwidth]{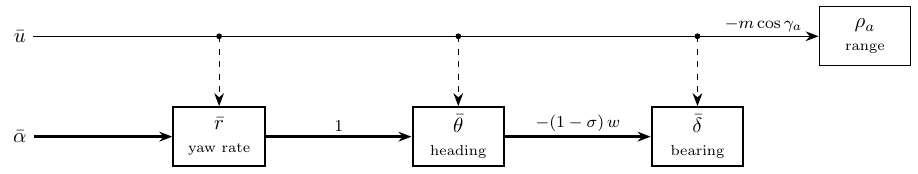}
\caption{The model \eqref{eq:znmodel}. Heavy black:
chain from the yaw moment to the bearing. Dashed: action of the surge
input.}
\label{fig:chain}
\end{figure*}

Figure~\ref{fig:chain} shows the model \eqref{eq:znmodel} as a block
diagram. The yaw moment acts
on the bearing through a chain of three states: the yaw
rate, the heading, and the bearing. Such a chain is what backstepping
requires. It is present in \eqref{eq:orig} too, but cannot be
used there: the action of
$\gamma$ on $\delta$ is divided by the range, and the sway speed,
driven by the yaw rate, acts on the range, the bearing and the
line-of-sight angle, outside the chain. In \eqref{eq:znmodel} neither
obstacle remains. By the normalization \eqref{eq:znvar}, the links of the
chain are bounded and do not vanish as $\rho_a\to0$.
By the zero-sideslip points \citep{pos}, the surge and sway speeds do not enter the chain or the range. The range is controlled directly by $\bar u$, but not separately,
since $\bar u$ enters every line of the chain, so speed and heading
are designed jointly, as in Theorem~\ref{thm:uni}.

\begin{assumption}\label{as:leader}
The leader's surge acceleration $\dot u_\e$ and yaw rate $r_\e$ are continuous, with bounded,
piecewise continuous derivatives. For all $t\ge0$,
\begin{equation}
\begin{gathered}
0<u_m\le u_\e(t)\le u_M,\quad|\dot u_\e(t)|\le a_M,\quad
|r_\e(t)|\le\omega_M,\\
\frac{X_M\,\omega_M^2}{Y_0^2}+\frac{|X_1|\,a_M}{|Y_0|}<u_m,\\
X_M=\max\bigl\{|X(u_m)|,\,|X(u_M)|\bigr\}.
\end{gathered}
\label{eq:omegaM}
\end{equation}
Its initial sway speed satisfies $|v_\e(0)|\le X_M\,\omega_M/|Y_0|$.
\end{assumption}

Under Assumption~\ref{as:leader}, the speed $w$ of the leader's zero-sideslip
point, given by \eqref{eq:w}, lies between
$u_{\min}=u_m-X_M\,\omega_M^2/Y_0^2-|X_1|\,a_M/|Y_0|>0$ and
$u_{\max}=u_M+X_M\,\omega_M^2/Y_0^2+|X_1|\,a_M/|Y_0|$.

\section{Underway docking with surge speed and yaw rate inputs (Theorem \ref{thm:kin})}\label{sec:kin}

\begin{theorem}[Exponential docking with surge speed and yaw rate inputs]\label{thm:kin}
Let Assumptions \ref{as:hull} and \ref{as:leader} hold. Let the follower
measure $\rho$, $\gamma$, $\theta$, $v_\p$, $u_\e$, $v_\e$, $r_\e$,
$\dot u_\e$, $\ddot u_\e$ and $\dot r_\e$. Consider \eqref{eq:z1}--\eqref{eq:z5} under
the dynamic feedback
\begin{subequations}\label{eq:klaw}
\begin{align}
\dot u_\p&=\frac{Y_0}{X_1}\Bigl(\sqrt{a^2+b^2}-u_\p\Bigr)-\frac{v_\p\,r_\p}{X_1},\label{eq:ku}\\
r_\p&=r^\ast,\label{eq:kr}
\end{align}
\end{subequations}
where
\begin{subequations}\label{eq:swhere}
\begin{align}
r^\ast&=r_\e(t)+\dot\delta_a-\frac{\dd}{\dd t}\arctan\frac ba\notag\\
&\quad+\Bigl(c-\frac{2(1-\sigma)\dot\rho_a}{\rho_a}\Bigr)z+m^2\chi ,\label{eq:rstar}\\
z&=\gamma_a-\arctan\frac ba,\label{eq:sz}\\
a&=w(t)+k_1m,\label{eq:sa}\\
b&=w(t)\sin\delta_a+\rho_a\,r_\e(t)-k_2\,m\,\delta_a\notag\\
&\quad-(1-\sigma)\,\delta_a\bigl[k_1m+w(t)(1-\cos\delta_a)\bigr],\label{eq:sb}\\
\chi&=\sigma\,m^2E_1+\frac{\delta_a}{\rho_a}\bigl(E_2-(1-\sigma)\,\delta_a\,E_1\bigr),\label{eq:schi}\\
E_1&=\frac{a(1-\cos z)+b\sin z}{z},\notag\\
E_2&=\frac{a\sin z-b(1-\cos z)}{z},\label{eq:sE}
\end{align}
\end{subequations}
with $\rho_a$, $\gamma_a$, $\delta_a$ given by \eqref{eq:zspdef}, $w$ by
\eqref{eq:w}, $m$ and $\sigma$ by \eqref{eq:zm}, \eqref{eq:zsig}, and $\dot\rho_a$,
$\dot\delta_a$ by the right sides of \eqref{eq:z1}, \eqref{eq:z2} with
$u_a=\sqrt{a^2+b^2}$, and where the gains $k_1$, $k_2$, $c$ are positive.
Let the normalized docking error be
\begin{equation}
\begin{aligned}
M&=\rho_a+|\bar\delta|+|\bar\theta-\bar\theta^\ast(\rho_a,t)|\\
&\quad+|u_\p-u_\e(t)|+|v_\p-v_\e(t)|,\\
\bar\theta^\ast(\rho_a,t)&=\frac{k_1\,\kappa_\e(t)}{(1-\sigma)\bigl(w(t)+k_1m\bigr)},
\end{aligned}
\label{eq:Mk}
\end{equation}
$\mu=\min\{Y_0/X_1,-Y_0\}$, $0<\nu<\min\{k_1,k_2,c,\mu\}$, and
\begin{align}
C_\rho&=\max\Bigl\{1+(1+D_\e)\frac{\omega_M}{u_{\min}},\ (1+D_\e)\,\ell^2,\notag\\
&\qquad\qquad\frac{|X_1|}{|Y_0|},\ \frac{1}{|Y_0|}\Bigr\},\notag\\
D_\e&=\frac{X_M}{|Y_0|}\sqrt{1+\frac{\omega_M^2}{Y_0^2}}\ .
\label{eq:Crho}
\end{align}
Then there are $\kappa,\kappa_r\in\mathcal K_\infty$
such that, for every initial condition with $\rho_a(0)>0$ and all $t\ge0$, the closed-loop
solution exists, is unique, and satisfies $\rho_a(t)>0$ and
\begin{subequations}\label{eq:kclaim}
\begin{align}
M(t)&\le\kappa\bigl(M(0)\bigr)\,e^{-\nu t},\label{eq:kclaim1}\\
\rho(t)+|\theta(t)|&\le C_\rho\,M(t),
\label{eq:kclaim2}\\
|r_\p(t)-r_\e(t)|&\le\kappa_r\bigl(M(0)\bigr)\,e^{-\nu t}.
\label{eq:kclaim3}
\end{align}
\end{subequations}
\end{theorem}

The proof is in Appendix~\ref{app:proof}.

\begin{remark}[Measurements]\label{rem:meas}
\rm The relative position and heading $\rho$, $\gamma$, $\theta$ can be
obtained by GNSS on both vessels in a moving-baseline configuration, by
range and line-of-sight sensors, or,
at close range, by lidar or cameras. The leader's speeds, yaw rate and
surge acceleration are transmitted from its navigation system by radio. The higher derivatives
$\ddot u_\e$ and $\dot r_\e$, and in Theorem~\ref{thm:dyn} also
$\dddot u_\e$ and $\ddot r_\e$, are not measured by standard sensors but
follow from the leader's guidance references or control signals, and
they vanish for a leader on a straight course at constant speed. For
AUVs under water, acoustic positioning and communication replace
GNSS and radio.
\end{remark}

\section{Underway docking with surge force and yaw moment inputs (Theorem \ref{thm:dyn})}\label{sec:dyn}

\begin{theorem}[Exponential docking with surge force and yaw moment inputs]\label{thm:dyn}
Let Assumptions \ref{as:hull} and \ref{as:leader} hold, let
$\ddot u_\e$ and $\dot r_\e$ be continuous, with bounded, piecewise
continuous derivatives, and $f_u$, $f_r$ continuously
differentiable. Let the follower measure $\rho$, $\gamma$, $\theta$, $u_\p$,
$v_\p$, $r_\p$, $u_\e$, $v_\e$, $r_\e$, $\dot u_\e$, $\ddot u_\e$,
$\dddot u_\e$, $\dot r_\e$ and $\ddot r_\e$. Consider
\eqref{eq:orig} under the feedback
\begin{subequations}\label{eq:dlaw}
\begin{align}
\tau_u&=-f_u+\frac{Y_0}{X_1}\Bigl(\sqrt{a^2+b^2}-u_\p\Bigr)-\frac{v_\p\,r_\p}{X_1},\label{eq:tauu}\\
\tau_r&=-f_r+\dot r^\ast-\Bigl(k_r-\frac{2(1-\sigma)\dot\rho_a}{\rho_a}\Bigr)(r_\p-r^\ast)+z,\label{eq:taur}
\end{align}
\end{subequations}
where $r^\ast$, $z$, $a$, $b$, $\sigma$, $\dot\rho_a$ are those of
Theorem~\ref{thm:kin}, $\dot r^\ast$ is the derivative of \eqref{eq:rstar}
along \eqref{eq:orig} with \eqref{eq:tauu}, and $k_1,k_2,c,k_r>0$. Let the normalized docking error be
\begin{subequations}\label{eq:Md}
\begin{align}
M&=\rho_a+|\bar\delta|+|\bar\theta-\bar\theta^\ast(\rho_a,t)|+|u_\p-u_\e(t)|\notag\\
&\quad+|v_\p-v_\e(t)|+|\bar r-\bar r^\ast(\rho_a,t)|,\label{eq:Md1}\\
\bar r^\ast(\rho_a,t)&=\bigl(\partial_t-k_1m\,\partial_{\rho_a}\bigr)\bar\theta^\ast(\rho_a,t)\notag\\
&\quad-2k_1(1-\sigma)^2\,\bar\theta^\ast(\rho_a,t),\label{eq:Md2}
\end{align}
\end{subequations}
with $\bar\theta^\ast$ as in \eqref{eq:Mk},
and let
$0<\nu<\min\{k_1,k_2,c,k_r,\mu\}$, with $\mu$ and $C_\rho$ as in Theorem~\ref{thm:kin}. Let
the leader's inputs be
$\tau_u^\e=\dot u_\e-f_u(u_\e,v_\e,r_\e)$ and
$\tau_r^\e=\dot r_\e-f_r(u_\e,v_\e,r_\e)$. Then there are
$\kappa,\kappa_\tau\in\mathcal K_\infty$ such that, for every
initial condition with $\rho_a(0)>0$ and all $t\ge0$, the closed-loop solution exists,
is unique, and satisfies $\rho_a(t)>0$ and
\begin{subequations}\label{eq:dclaim}
\begin{align}
M(t)&\le\kappa\bigl(M(0)\bigr)\,e^{-\nu t},\label{eq:dclaim1}\\
\rho(t)+|\theta(t)|&\le C_\rho\,M(t),
\label{eq:dclaim2}\\
|\tau_u(t)-\tau_u^\e(t)|+|\tau_r(t)-\tau_r^\e(t)|&\le\kappa_\tau\bigl(M(0)\bigr)\,e^{-\nu t}.
\label{eq:dclaim3}
\end{align}
\end{subequations}
\end{theorem}

\begin{figure*}[t]
\centering
\includegraphics[width=0.85\textwidth]{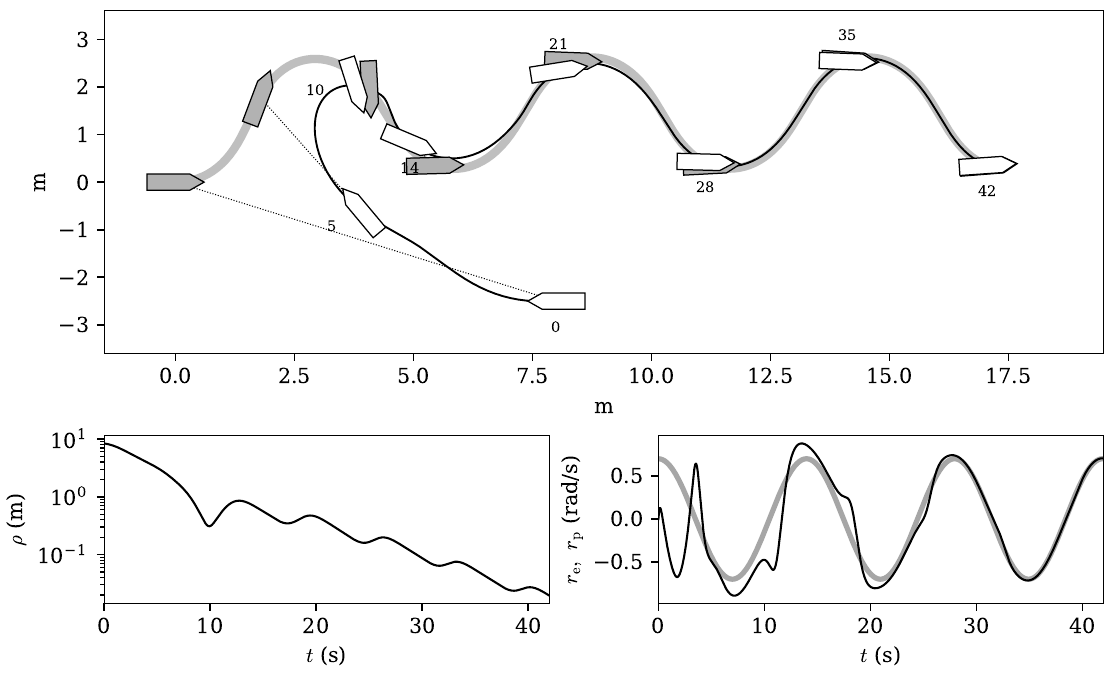}
\caption{Underway docking under the feedback \eqref{eq:dlaw} of
Theorem~\ref{thm:dyn} to a leader (grey) in a slalom, with
$u_\e=0.5$~m/s, $r_\e(t)=0.7\cos(0.45t)$~rad/s. Top: paths, with the hulls at the marked
times (s). Bottom: distance between the hulls, and the yaw rates of
the leader (grey) and the follower (black). Gains $k_1=0.15$, $k_2=0.6$,
$c=1$, $k_r=2$, $\ell=2$~m.}
\label{fig:slalom}
\end{figure*}

Figure~\ref{fig:slalom} shows the feedback \eqref{eq:dlaw} for the Cybership~I model of
\citet{pos} and a leader in a slalom, whose heading swings by
$\pm89^\circ$ and whose sideslip angle reaches $37^\circ$. The follower starts at rest with the opposite heading, turns around, enters the leader's path from astern, and docks
while the leader turns.

The universal controller of \citet{djp} tracks a reference ship and
stabilizes it at a fixed pose with one control law, in Cartesian error
coordinates, by adding the periodic term $\lambda_2\cos(\lambda_3t)$ to
the gain of its heading reference. A slow leader on a
straight course is the case of underway docking closest to stabilization
at a fixed pose, and it separates the two control laws.

\begin{figure}[!t]
\centering
\includegraphics[width=0.85\columnwidth]{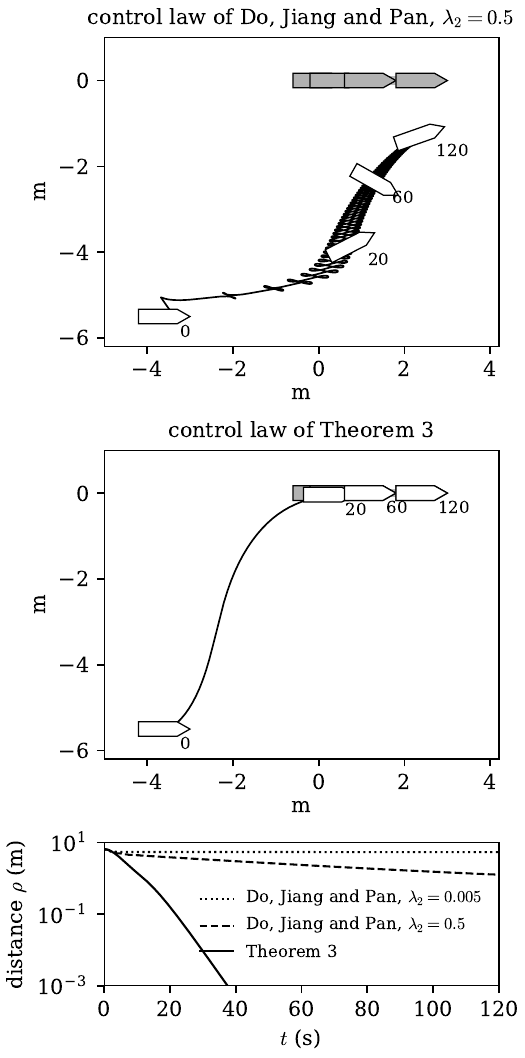}
\caption{Docking to a leader (grey) on a straight course at
$u_\e=0.02$~m/s, under the control laws of \citet{djp} and of
Theorem~\ref{thm:dyn}. Top and middle: paths, with the hulls at $t=0$, $20$,
$60$, $120$~s. Bottom: distance between the hulls. Gains of \citet{djp}:
$\lambda_1=1.5$, $\lambda_3=3.5$, $k_1=0.44$, $k_2=0.01$, $k_3=0.8$,
$c_1=c_2=5$; of Theorem~\ref{thm:dyn}: $k_1=0.3$, $k_2=0.6$, $c=1$,
$k_r=2$, $\ell=2$~m.}
\label{fig:cmp}
\end{figure}

Figure~\ref{fig:cmp} shows a follower that starts abeam and astern of a
slow leader, for the Cybership~I model with
$d_{23}=d_{32}=0$, so that the inertia and damping matrices are diagonal
as in \citet{djp}. Under the control law of
\citet{djp} with $\lambda_2=0.005$, the value used there, the
follower keeps its lateral offset. With $\lambda_2=0.5$ it closes by
zigzagging at the frequency $\lambda_3$, and the distance decays slowly.
Under the control law of Theorem~\ref{thm:dyn} the follower closes on a smooth
path and arrives from astern, and the distance decays at the rate $k_1$.

\section{Arrival at turning leader}\label{sec:arrival}

Theorems \ref{thm:kin} and \ref{thm:dyn} give the decay of the distance
$\rho_a$ between the zero-sideslip points. Parallel arrival is a statement
about the heading error $\theta$ per distance $\rho$ between the hulls. The
two distances differ through the speed errors $u_\p-u_\e$ and
$v_\p-v_\e$, because each hull is displaced from its zero-sideslip point
in proportion to its speeds. The speed errors per unit of distance,
$(u_\p-u_\e)/\rho_a$ and $(v_\p-v_\e)/\rho_a$, obey the hull's equations of
the speed errors, divided by $\rho_a$. With each coefficient of these
equations replaced by its limit as the follower closes, the equations are
\begin{subequations}\label{eq:UV}
\begin{align}
\dot U&=-\Bigl(\frac{Y_0}{X_1}-k_1\Bigr)U-\frac{r_\e(t)}{X_1}\,V+\frac{Y_0}{X_1}\,k_1-\frac{v_\e(t)}{X_1}\,g(t),\label{eq:Ueq}\\
\dot V&=(Y_0+k_1)\,V+X_1\,r_\e(t)\,U+X(u_\e(t))\,g(t),\label{eq:Veq}\\
g(t)&=k_1\,\kappa_\e(t)-\dot\kappa_\e(t).\label{eq:g}
\end{align}
\end{subequations}
where $k_1$ is the rate at which $\rho_a$ closes and $g$ is the limit of
the yaw-rate error per unit of distance, $(r_\p-r_\e)/\rho_a$. We denote
the solution of \eqref{eq:UV} by $(U,V)$, and prove next that the
speed errors per unit of distance converge to it.

\begin{proposition}[Parallel arrival at turning leader]\label{prop:par}
Let the hypotheses of Theorem \ref{thm:kin} or of Theorem \ref{thm:dyn}
hold, let $(U,V)$ be a solution of \eqref{eq:UV}, and let $P(t)\ge0$
be given by
\begin{equation}
\begin{aligned}
P^2(t)&=\Bigl(\frac{Y_0u_\e(t)+X_1\dot u_\e(t)}{Y_0\,w(t)}+\frac{X_1}{Y_0}\,U(t)\Bigr)^2\\
&\quad+\Bigl(\frac{X(u_\e(t))\,\kappa_\e(t)+V(t)}{Y_0}\Bigr)^2 .
\end{aligned}
\label{eq:P}
\end{equation}
Then, as $t\to\infty$,
\begin{equation}
\frac{\theta(t)}{\rho_a(t)}+\kappa_\e(t)\to0 .
\label{eq:slopea}
\end{equation}
If $k_1<\mu$, then $(u_\p-u_\e)/\rho_a-U\to0$, $(v_\p-v_\e)/\rho_a-V\to0$
and $\rho/\rho_a-P\to0$. If also
$\liminf_{t\to\infty}P(t)>0$, then
\begin{equation}
\frac{\theta(t)}{\rho(t)}+\frac{\kappa_\e(t)}{P(t)}\to0 .
\label{eq:slope}
\end{equation}
\end{proposition}

\begin{figure}[t]
\centering
\includegraphics[width=0.85\columnwidth]{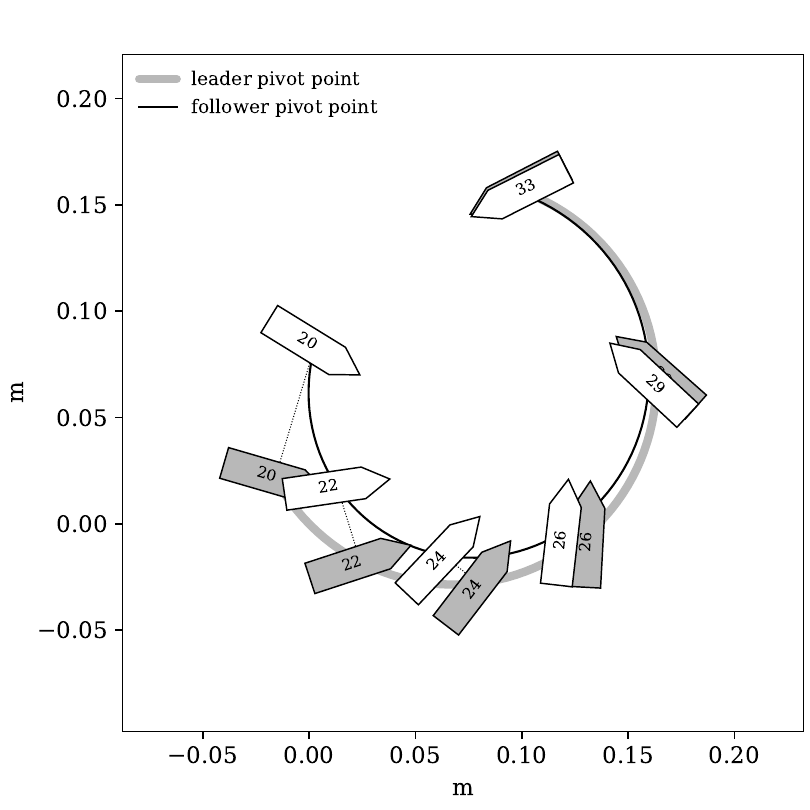}
\caption{Arrival of the follower (white) at a leader (grey) that turns at
$r_\e=0.3$~rad/s with $u_\e=0.02$~m/s, under the feedback of
Theorem~\ref{thm:dyn}; times in seconds. The
heading error decays in proportion to the separation
(Proposition~\ref{prop:par}).}
\label{fig:arrival}
\end{figure}

The proof is in Appendix~\ref{app:par}. By \eqref{eq:slopea}, the follower's heading is
the direction of the leader's path where the follower is, so it differs
from the leader's heading only by the turn that the leader makes over the
distance that remains. By \eqref{eq:slope}, in the distance between the
hulls, $\theta\approx-\kappa_\e\,\rho/P$. Figure~\ref{fig:arrival} shows
the arrival at a leader that turns at a low surge speed, for which the
curvature $\kappa_\e$ is large.

\section{Underway docking with offset and different hull}\label{sec:vl}

Let the follower dock at a point fixed to the leader, at the distance
$d\ge0$ from the leader's pivot point and at the angle $\beta$ from the
leader's heading: astern for $\beta=\pi$, abeam for $\beta=\pm\pi/2$. This
offset point moves along and across the leader's heading at the speeds
\begin{equation}
u_o=u_\e-r_\e\,d\sin\beta,\qquad v_o=v_\e+r_\e\,d\cos\beta .
\label{eq:off}
\end{equation}
Let the leader's hull also differ from the follower's, as when a small autonomous surface vessel (ASV) docks with a larger vessel, or an AUV with a surface vessel. The leader's sway speed
then obeys, in place of \eqref{eq:leader},
\begin{equation}
\begin{gathered}
\dot v_\e=Y_\e v_\e+B_\e r_\e(t),\qquad|B_\e|\le B_M,\\
|v_\e(0)|\le b_\e\,\omega_M,\qquad b_\e=\frac{B_M}{|Y_\e|},
\end{gathered}
\label{eq:vl-sway}
\end{equation}
with $Y_\e<0$ and $B_\e=X_{\e0}+X_{\e1}u_\e(t)$, where the coefficients
$X_{\e0}$, $X_{\e1}$ are known to the follower.

\begin{figure}[t]
\centering
\includegraphics[width=0.85\columnwidth]{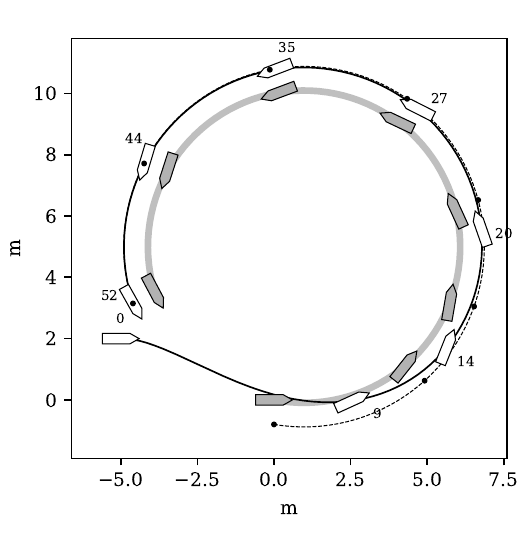}
\caption{Offset docking abeam of a leader in a circle, under the
feedback \eqref{eq:dlaw} of Theorem~\ref{thm:dyn} applied to the virtual
leader as in Theorem~\ref{thm:vl}, for equal hulls, $d=0.8$~m,
$\beta=-\pi/2$, $u_\e=0.5$~m/s and $r_\e=0.1$~rad/s. Paths of the leader
(grey), the offset point (dashed) and the follower (solid), with the
hulls and the offset point (dot) at the marked times (s). Gains
as in Figure~\ref{fig:slalom}.}
\label{fig:offset}
\end{figure}

\begin{definition}[Virtual leader]\label{def:vl}
The virtual leader is the vessel with the hull coefficients of the
follower, with its pivot point at the offset point, and with the
heading $\psi_\e+\eta$, where $\psi_\e$ is the leader's heading and
\begin{subequations}\label{eq:vl}
\begin{align}
\dot\eta&=\frac{\dot v_o\cos\eta-\dot u_o\sin\eta-Y_0v_d-X(u_d)\,r_\e}
{u_d+X(u_d)},\notag\\
\eta(0)&=0,\label{eq:vl-gen}\\
u_d&=u_o\cos\eta+v_o\sin\eta,\label{eq:vl-ud}\\
v_d&=-u_o\sin\eta+v_o\cos\eta ,\label{eq:vl-vd}
\end{align}
\end{subequations}
with the derivatives of \eqref{eq:off} taken along
\eqref{eq:vl-sway}. Its yaw rate and the forward speed of its
zero-sideslip point are
\begin{subequations}\label{eq:vl-rw}
\begin{align}
r_d&=r_\e+\dot\eta,\label{eq:vl-rd}\\
w_d&=u_d+\frac{X_1}{Y_0}\bigl[\dot u_o\cos\eta+\dot v_o\sin\eta
+v_d\,\dot\eta\bigr]+\frac{v_d\,r_d}{Y_0}.\label{eq:vl-wd}
\end{align}
\end{subequations}
\end{definition}

The virtual leader moves as the follower's hull would move along the
path of the offset point: $u_d$, $v_d$ are the components of the
velocity of the offset point along and across the heading
$\psi_\e+\eta$, and $\dot v_d=Y_0v_d+X(u_d)\,r_d$. The heading of a follower that stays at the offset point differs
from the leader's heading by a solution $\eta$ of the differential
equation in \eqref{eq:vl-gen}. For equal hulls, $\eta\equiv0$ if and
only if, for all $t\ge0$,
\begin{equation}
d\,\bigl[(\dot r_\e-Y_0\,r_\e)\cos\beta+X_1\,r_\e^2\sin\beta\bigr]=0,
\label{eq:off-cond}
\end{equation}
which holds without offset and on a straight course.

\begin{lemma}[Bounds on virtual leader]\label{lem:vl}
Let $Y_0<0$ and $X_1<0$, let $u_o$ be continuously differentiable
with $0<u_m\le u_o(t)\le u_M$ and $|\dot u_o(t)|\le a_M$, and
let \eqref{eq:vl-sway} hold with $r_\e$ piecewise continuous and
$|r_\e(t)|\le\omega_M\le\omega^\ast$, and, if $d>0$, with $r_\e$
continuously differentiable and $|\dot r_\e(t)|\le|Y_0|\,\omega_M$, where
\begin{subequations}\label{eq:vl-om}
\begin{align}
\omega^\ast&=\sqrt{\frac{w_m\,d_m}{2\bigl(d_m\,K_w+w_m|1+X_1|K_u\bigr)}},
\label{eq:vl-om1}\\
h_0&=|Y_0|\,u_m-a_M>0,\notag\\
w_m&=u_m-\frac{|X_1|\,a_M}{|Y_0|}>0,\notag\\
d_m&=\min\bigl\{u_m+X(u_m),\,u_M+X(u_M)\bigr\}>0,
\label{eq:vl-h}\\
b_o&{}=b_\e+d\,|\cos\beta|,\notag\\
B_o&{}=B_M+\tfrac12|Y_0|\,d\,|\cos\beta|,\notag\\
C_0&=|Y_\e-Y_0|\,b_\e+B_M+X_M+\frac{|X_1|\,u_m}{2}\notag\\
&\quad{}+2\,|Y_0|\,d\,|\cos\beta|,\notag\\
K_\eta&=\frac{2C_0}{h_0},\notag\\
K_r&=\frac{2C_0}{d_m}\Bigl(1+\frac{2(|Y_0|\,u_M+a_M)}{h_0}\Bigr),\notag\\
K_u&=\frac{u_M\,K_\eta^2}{2}+b_oK_\eta,
\label{eq:vl-C}\\
K_w&=K_u+\frac{|X_1|\bigl(a_MK_\eta^2/2+2B_oK_\eta\bigr)}{|Y_0|}\notag\\
&\quad+\frac{(u_M\,K_\eta+b_o)\bigl(1+|1+X_1|K_r\bigr)}{|Y_0|}.
\label{eq:vl-K}
\end{align}
\end{subequations}
Then $\eta$ exists for all $t\ge0$, and for all $t\ge0$
\begin{gather}
|\eta|\le K_\eta\,\omega_M,\qquad u_d+X(u_d)\ge\frac{d_m}2,\notag\\
|r_d|\le(1+K_r)\,\omega_M,\notag\\
\frac{w_m}2\le w_d\le u_M+\frac{|X_1|\,a_M}{|Y_0|}+\frac{w_m}2 .
\label{eq:vl-b}
\end{gather}
\end{lemma}

\begin{theorem}[Exponential offset docking to leader with different hull]\label{thm:vl}
Let the hypotheses of Lemma~\ref{lem:vl} hold, and let $\dot u_\e$ and
$r_\e$ be continuous, with bounded, piecewise continuous
derivatives, and likewise $\dot r_\e$ if $d>0$. Let $\rho$ and
$\gamma$ be the distance and the line-of-sight angle from the follower's
pivot point to the offset point. Let
the follower measure $\rho$, $\gamma$, $\theta$, $v_\p$, $u_\e$, $v_\e$,
$r_\e$, $\dot u_\e$, $\ddot u_\e$ and $\dot r_\e$, and $\ddot r_\e$
if $d>0$, and apply the feedback
\eqref{eq:klaw} of Theorem~\ref{thm:kin} with
\begin{equation}
\begin{gathered}
(\psi_\e,\,u_\e,\,v_\e,\,r_\e,\,w,\,\theta)\ \text{ replaced by }\\
(\psi_\e+\eta,\,u_d,\,v_d,\,r_d,\,w_d,\,\theta-\eta).
\end{gathered}
\label{eq:vl-rule}
\end{equation}
Let the surge force and the yaw moment of the virtual leader be
$\tau_u^d=\dot u_d-f_u(u_d,v_d,r_d)$ and
$\tau_r^d=\dot r_d-f_r(u_d,v_d,r_d)$. Then the estimates
\eqref{eq:kclaim1}, \eqref{eq:kclaim3} hold with the replacement
\eqref{eq:vl-rule}, the distance $\rho$ from the follower's hull to
the offset point and the heading
error $\theta-\eta$ decay exponentially, and
$|\eta(t)|\le K_\eta\,\omega_M$ for all $t\ge0$. The same holds for the
feedback \eqref{eq:dlaw} and the estimates \eqref{eq:dclaim1},
\eqref{eq:dclaim3} of Theorem~\ref{thm:dyn}, with
$(\tau_u^\e,\tau_r^\e)$ replaced by $(\tau_u^d,\tau_r^d)$, if
$\ddot u_\e$ and $\dot r_\e$ are continuous, with bounded, piecewise
continuous derivatives, and likewise $\ddot r_\e$ if $d>0$, $f_u$, $f_r$ are continuously differentiable,
and the follower also measures $u_\p$, $r_\p$, $\dddot u_\e$ and
$\ddot r_\e$, and $\dddot r_\e$ if $d>0$.
\end{theorem}

The proof is in Appendix~\ref{app:vl}.

In Figure~\ref{fig:offset} the follower docks on the starboard
side of a leader in a circle. It starts on the port side, crosses astern of the leader, reaches the offset point from astern, and stays abeam of the leader around the circle. Its
heading converges to $\psi_\e+\eta$, with $\eta$ converging to a constant.

\section{Conclusions}

In order to solve the problem of underway docking for underactuated
ships, under the challenge of sideslip and no sway actuation, we have had
to build the design techniques starting from unicycle, for which underway
docking with bearing regulation, namely, from astern, had never been
solved. And then provide extensions from unicycle
(Theorem~\ref{thm:uni}) to a kinematic ship (Theorem~\ref{thm:kin}), and
from kinematic ship to a dynamic underactuated ship
(Theorem~\ref{thm:dyn}).
Theorem~\ref{thm:vl} moves the docking point astern or abeam of the
leader, where a smaller vessel is picked up. We found that a
follower without sway actuation holds it behind a turning leader
only with a heading different from the leader's, by an angle we
bound.

The ideas of this paper come from five lines of work. Each of them had
to be advanced for a leader that keeps moving, and these advances are
what we contribute.

\vspace*{-15pt}\paragraph{Stabilization at a fixed pose.}
From \citet{pos} we take the zero-sideslip point, at which a ship with
sway moves as a unicycle, and the energy of the surge and sway speed
errors, now for both vessels. The mechanism of polar parking does not
carry over. There the speed is proportional to the range, the division by
the range cancels, and the angles need only converge. Behind a moving
leader the speed stays near the leader's, the division remains, and the
angles have to decay faster than the range. Docking under way is a
different problem, not parking with a moving target.

\vspace*{-15pt}\paragraph{Pursuit.}
The polar model of two vehicles and the objective of arrival along the
other vehicle's heading are those of pursuit \citep{dpe}, and the speed
control law that is defined for every bearing is that of \citet{wk}.
Pursuit and docking differ in how they end. A pursuer is faster than the
evader and captures it in finite time, with a steering gain divided by
the range. A follower that docks settles at the leader's speed and stays at zero
range for all time, where the division by the range acts. We moved the division from the gain into the states. The result is
exponential stability of the leader's trajectory, with a surge force and
a yaw moment that converge to the leader's.

\vspace*{-15pt}\paragraph{Range normalization.}
Angles divided by the range gave exponential parking of the bicycle
and the tractor-trailer \citep{bike,tt}. For a moving leader,
we carry the division further in two ways. First, the bearing is divided
by the range, and the heading error, taken relative to the leader's path,
by its square. The
normalized bearing is then the curvature of the arc on which the follower
reaches the leader along the leader's heading, and the normalized model
\eqref{eq:znmodel} has no division by the range. Second, we divide by a weight
that is the range near the leader and a constant far from it, and the
range enters the Lyapunov function through the integral of this
weight. The closing
speed is then bounded at any distance, and the
normalization brings no powers of the range into the yaw rate. The
normalized model precedes any
control law: it is the model of underway docking, as the polar model is the
model of parking.

\vspace*{-15pt}\paragraph{Nonmodular backstepping.}
The control law is designed by the nonmodular backstepping of
\citet{part2}, whose first instance is in \citet{restrepo}: the speed and
the commanded heading are
chosen jointly, for the range and the bearing together, and the yaw rate
then drives the heading to its commanded value. In the modular designs of
\citet{part1}, the speed is chosen first and the steering is
designed for the angles alone. Our virtual control differs from those of
\citet{part2} and \citet{wk} because the problems differ. In parking, the commanded
velocity vanishes with the range. In pursuit, the commanded
velocity along the line of sight exceeds the evader's speed by a fixed
fraction. Under way, it contains the leader's speed and turning, it is designed for the
bearing divided by the range, and the backstepping error is the heading
error divided by the range squared.

\vspace*{-15pt}\paragraph{Tracking in Cartesian coordinates.}
The control law of \citet{djp} both tracks a moving ship
and stabilizes it at a fixed pose. The
time-varying term that parking in Cartesian coordinates requires stays in
that control law when the leader moves, and behind a slow leader the follower
zigzags where the control law of Theorem~\ref{thm:dyn} approaches smoothly
from astern (Figure~\ref{fig:cmp}). Cartesian error
coordinates also have no quantity for the direction of approach, which in
the normalized states is the bearing divided by the range. Two dedicated
control laws, that of
\citet{pos} for a target at rest and that of Theorem~\ref{thm:dyn} for a
target that moves, solve the two problems better than one.

\appendix

\section{Proof of Theorems \ref{thm:uni}, \ref{thm:kin} and \ref{thm:dyn}}\label{app:proof}

The proof has four steps. The backstepping errors, divided by the same
powers of the range as the states that they replace, obey a target system
(Section~\ref{app:A1}) with a strict Lyapunov function
(Section~\ref{app:A2}). The surge and sway speed errors are a cascade
driven by that system (Section~\ref{app:A3}). The estimate passes from the backstepping errors to the tracking errors of $M$
(Section~\ref{app:A4}).

The proof is given for Theorem~\ref{thm:dyn}. Theorem~\ref{thm:kin} is the
case $\bzeta\equiv0$. Theorem~\ref{thm:uni} is the case $\bzeta\equiv0$
without the speed errors, with
$(\rho,\delta,\gamma,w_\e,\omega_\e,w_\p,\omega_\p)$ in place of
$(\rho_a,\delta_a,\gamma_a,w,r_\e,u_a,r_\p)$. The angles $\gamma_a$,
$\theta$ and $\delta_a=\gamma_a+\theta$ are continued continuously in time. The closed loop is studied in the state
$(\rho_a,\delta_a,\gamma_a,u_\p,v_\p,r_\p)$ of \eqref{eq:zmodel}, on
$\rho_a>0$, where its right side is locally Lipschitz; a passage of $\rho$
through zero does not interrupt the solution.

\subsection{Normalized backstepping errors and target system (Theorems \ref{thm:uni}, \ref{thm:kin}, \ref{thm:dyn})}\label{app:A1}

The commanded velocity $(a,b)$ and the angle $z$ from it to the
follower's heading are used as in \citet{wk}. The new step is that the
backstepping errors are divided by the same powers of $m$ as the states
that they replace: $\bz=z/m^2$ and $\bzeta=(r_\p-r^\ast)/m^2$.
A normalized variable has, in its derivative, the derivative of its
weight. With $\dot m/m=(1-\sigma)\dot\rho_a/\rho_a$,
\begin{equation}
\begin{aligned}
\frac{\dd}{\dd t}\frac{\delta_a}{m}&=\frac{\dot\delta_a}{m}-\frac{(1-\sigma)\dot\rho_a}{\rho_a}\,\bd,\\
\frac{\dd}{\dd t}\frac{z}{m^2}&=\frac{\dot z}{m^2}-\frac{2(1-\sigma)\dot\rho_a}{\rho_a}\,\bz .
\end{aligned}
\label{eq:A-w}
\end{equation}
The feedback cancels these terms. The last term of $b$ in \eqref{eq:sb}
cancels the term with $\bd$ at $\bz=0$. The gain
$-2(1-\sigma)\dot\rho_a/\rho_a$ in \eqref{eq:rstar} cancels the term with
$\bz$, and the same gain in \eqref{eq:taur} cancels its counterpart for
$\bzeta$. The target system therefore has no division by the range.

By \eqref{eq:z4} with \eqref{eq:ku} or \eqref{eq:tauu},
$u_a=\sqrt{a^2+b^2}$. With $\gamma_a=\arctan(b/a)+z$ in \eqref{eq:z1},
\eqref{eq:z2},
$\dot\rho_a=w\cos\delta_a-a+zE_1$ and
$\rho_a\dot\delta_a=b-w\sin\delta_a-\rho_ar_\e+zE_2$.
The functions $E_1$, $E_2$ are the exact remainders when the heading
differs from the commanded direction by $z$. Substituting \eqref{eq:sa},
\eqref{eq:sb}, $\delta_a=m\bd$ and $z=m^2\bz$, and using \eqref{eq:A-w},
gives the first two lines of the target system
\begin{subequations}\label{eq:target}
\begin{align}
\dot\rho_a&=-k_1m-w(t)\bigl(1-\cos(m\bd)\bigr)+m^2\,\bz\,E_1,\label{eq:t1}\\
\dot{\bd}&=-k_2(1-\sigma)\,\bd+\bz\,(1-\sigma)\bigl(E_2-(1-\sigma)\,m\,\bd\,E_1\bigr),\label{eq:t2}\\
\dot{\bz}&=-c\,\bz-\chi-\bzeta,\label{eq:t3}\\
\dot{\tu}&=-\frac{Y_0}{X_1}\,\tu+\frac{Y_0}{X_1}\,(u_a-w(t))\notag\\
&\quad-\frac{\tv\,r_\p+v_\e(t)\,(r_\p-r_\e(t))}{X_1},\label{eq:t4}\\
\dot{\tv}&=Y_0\,\tv+X_1\,\tu\,r_\p+X(u_\e(t))\,(r_\p-r_\e(t)),\label{eq:t5}\\
\dot{\bzeta}&=-k_r\,\bzeta+\bz ,\label{eq:t6}
\end{align}
\end{subequations}
with the speed errors $\tu=u_\p-u_\e$ and $\tv=v_\p-v_\e$. The third
line follows from
$\dot z=\dot\delta_a+r_\e-r_\p-\frac{\dd}{\dd t}\arctan(b/a)$ and
\eqref{eq:rstar}, which give
$\dot z=\bigl(2(1-\sigma)\dot\rho_a/\rho_a-c\bigr)z-m^2\chi-(r_\p-r^\ast)$,
and the sixth likewise from \eqref{eq:taur}. The fourth and fifth
are \eqref{eq:z4}, \eqref{eq:z5} minus \eqref{eq:leader} and
$\dot u_\e=(Y_0/X_1)(w-u_\e)-v_\e r_\e/X_1$, which is \eqref{eq:w}. With
\eqref{eq:kr}, $\bzeta\equiv0$.

\subsection{Lyapunov function and decay rate (Theorems \ref{thm:uni}, \ref{thm:kin}, \ref{thm:dyn})}\label{app:A2}

The range enters the Lyapunov function through a function $\Phi$ with
$\Phi'=\sigma$, which is quadratic near the leader and linear far from
the leader:
\begin{equation}
\begin{aligned}
V&=\Phi(\rho_a)+\tfrac12\bigl(\bd^2+\bz^2+\bzeta^2\bigr),\\
\Phi(\rho_a)&=\rho_a-\ell\ln\Bigl(1+\frac{\rho_a}{\ell}\Bigr).
\end{aligned}
\label{eq:A-V}
\end{equation}
With this choice the cross terms of $\bz$ are those that $\chi$ in
\eqref{eq:schi} cancels, and $\dot V$ is an identity. The
term $\sigma m^2\bz E_1$ from $\Phi'\dot\rho_a$ and the term
$\bd\,\bz(1-\sigma)(E_2-(1-\sigma)m\bd E_1)$ from $\bd\,\dot{\bd}$ are
cancelled by $-\bz\chi$, and $-\bz\bzeta$ is cancelled by $\bzeta\bz$:
\begin{equation}
\dot V=-\sigma\bigl[k_1m+w(1-\cos\delta_a)\bigr]-k_2(1-\sigma)\,\bd^2-c\,\bz^2-k_r\,\bzeta^2 .
\label{eq:Vdot}
\end{equation}
Let $\lambda=\min\{k_1,k_2,c,k_r\}$. We write \eqref{eq:Vdot} as
\begin{equation}
\begin{gathered}
\dot V=-\Pi_1-\Pi_2-\Pi_3-\Pi_4-\Pi_5,\\
\Pi_1=\sigma\,k_1m,\qquad\Pi_2=\sigma\,w\,(1-\cos\delta_a),\\
\Pi_3=k_2(1-\sigma)\,\bd^2,\quad\ \Pi_4=c\,\bz^2,\quad\ \Pi_5=k_r\,\bzeta^2,
\end{gathered}
\label{eq:A-Pi}
\end{equation}
and bound each term from below. Since
$x-x^2/2\le\ln(1+x)\le x-x^2/(2+2x)$ for $x\ge0$, at $x=\rho_a/\ell$,
\begin{equation}
\frac{\rho_a^2}{2(\ell+\rho_a)}\le\Phi(\rho_a)\le\frac{\rho_a^2}{2\ell}.
\label{eq:A-Phi}
\end{equation}
By
$m=\ell\sigma$, $(1-\sigma)\rho_a=\ell\sigma$ and \eqref{eq:A-Phi},
\begin{equation}
\begin{aligned}
\Pi_1&=k_1\,\ell\,\sigma^2=k_1(1-\sigma)^2\,\frac{\rho_a^2}{\ell}\\
&\ge2\lambda\,(1-\sigma)^2\,\Phi(\rho_a).
\end{aligned}
\label{eq:A-P1}
\end{equation}
Since $w>0$ and $0\le\sigma<1$,
\begin{equation}
\begin{gathered}
\Pi_2\ge0,\qquad\Pi_3\ge\lambda\,(1-\sigma)^2\,\bd^2,\\
\Pi_4\ge\lambda\,(1-\sigma)^2\,\bz^2,\qquad\Pi_5\ge\lambda\,(1-\sigma)^2\,\bzeta^2,
\end{gathered}
\label{eq:A-P25}
\end{equation}
where the bound on $\Pi_3$ uses $1-\sigma\ge(1-\sigma)^2$. By
\eqref{eq:A-Pi}, \eqref{eq:A-P1}, \eqref{eq:A-P25} and \eqref{eq:A-V},
\begin{equation}
\dot V\le-2\lambda\,(1-\sigma)^2\,V .
\label{eq:A-Vineq}
\end{equation}
Hence $V$ is nonincreasing, $\rho_a(t)\le\Phi^{-1}(V(0))$, and $\bd$,
$\bz$, $\bzeta$ are bounded. By \eqref{eq:t1}, $\dot\rho_a\ge-C\rho_a$
with $C$ depending on $V(0)$, so $\rho_a(t)>0$ and the
solution exists for all $t\ge0$.

The factor $(1-\sigma)^2$ in \eqref{eq:A-Vineq} is small far from the
leader. There the follower closes at the speed $k_1m$, which is at most
$k_1\ell$, and the decay is not exponential. The decay is exponential
once the follower is near the leader, and the overshoot function accounts
for the time of the approach. Let $\nu<\lambda$, let $\rho_\nu>0$ be
defined by $(1-\sigma(\rho_\nu))^2=\nu/\lambda$, and let
$T(s)=(\Phi^{-1}(s)+\ell)^2\ln^{+}\!\bigl(s/\Phi(\rho_\nu)\bigr)/(2\lambda\,\ell^2)$.
By \eqref{eq:A-Vineq}, $V$ reaches $\Phi(\rho_\nu)$ no later than
$T(V(0))$ and satisfies $\dot V\le-2\nu V$ thereafter, so
$V(t)\le V(0)\,e^{2\nu T(V(0))}e^{-2\nu t}$. With
$\Theta=(\rho_a,\bd,\bz,\bzeta)$ and $A(s)=\Phi(s)+s^2/2$,
\begin{equation}
\begin{gathered}
|\Theta(t)|\le\kappa_1\bigl(|\Theta(0)|\bigr)\,e^{-\nu t},\\
\kappa_1(s)=\sqrt{2\max\{\ell+\Phi^{-1}(A(s)),1\}\,A(s)}\ e^{\nu T(A(s))}.
\end{gathered}
\label{eq:A-k1}
\end{equation}

\subsection{Speed errors as a cascade (Theorems \ref{thm:kin}, \ref{thm:dyn})}\label{app:A3}

The energy $W$ of the surge and sway speed errors is that of \citet{pos}.
In \eqref{eq:t4}, \eqref{eq:t5} the speed errors are driven by the
docking error $\Theta$, and their estimate follows from the estimate of
$\Theta$. For $W=\tfrac12(X_1^2\tu^2+\tv^2)=\tfrac12s^2$, the terms with
$\tu\,\tv\,r_\p$ cancel:
$\dot W=-X_1Y_0\,\tu^2+Y_0\,\tv^2+X_1Y_0\,\tu\,(u_a-w)
+\bigl(X(u_\e)\,\tv-X_1v_\e\,\tu\bigr)(r_\p-r_\e)\le-\mu\,s^2+s\,g$,
with $g=|Y_0|\,|u_a-w|+(V_\e+X_M)\,|r_\p-r_\e|$ and
$V_\e=X_M\,\omega_M/|Y_0|$. The functions
$u_a-w=\sqrt{a^2+b^2}-w$ and $r_\p-r_\e=m^2\bzeta+(r^\ast-r_\e)$ are smooth
in $\Theta$ for $\rho_a\ge0$, vanish at $\Theta=0$, and depend on $t$
through the bounded signals $w$, $\dot w$, $r_\e$, $\dot r_\e$. Hence
$g\le L_0(|\Theta|)\,|\Theta|$ with $L_0$ continuous and nondecreasing,
and, for $\nu<\mu$, with $\kappa_1=\kappa_1(|\Theta(0)|)$,
\begin{equation}
s(t)\le\Bigl(s(0)+\frac{L_0(\kappa_1)\,\kappa_1}{\mu-\nu}\Bigr)e^{-\nu t}.
\label{eq:A-s}
\end{equation}

\subsection{From backstepping errors to tracking errors (Theorems \ref{thm:uni}, \ref{thm:kin}, \ref{thm:dyn})}\label{app:A4}

The estimates \eqref{eq:A-k1}, \eqref{eq:A-s} contain the backstepping
errors $\bz$ and $\bzeta$, which are not errors relative to the leader's
trajectory. The tracking errors of $M$ are
\begin{equation}
\begin{gathered}
\tilde\theta=\bar\theta-\bar\theta^\ast=\frac{\theta+\vartheta}{m^2},\qquad
\tilde r=\frac{r_\p-r_{\rm ref}}{m^2},\\
\vartheta(\rho_a,t)=\frac{\rho_a\,r_\e(t)}{w(t)+k_1m},
\end{gathered}
\label{eq:A-err}
\end{equation}
with $r_{\rm ref}=r_\e-(\partial_t-k_1m\,\partial_{\rho_a})\vartheta$.
They differ from $\bz$ and $\bzeta$ by a heading offset $h$ and a
yaw-rate offset $G$,
\begin{equation}
\begin{gathered}
\bz=-\tilde\theta-h,\qquad h=\frac{\arctan(b/a)-m\bd-\vartheta}{m^2},\\
\bzeta=\tilde r-G,\qquad G=\frac{r^\ast-r_{\rm ref}}{m^2}.
\end{gathered}
\label{eq:A-hG}
\end{equation}
The two offsets vanish at the leader, as shown next, so the estimate
passes to the tracking errors.

\emph{Heading offset.} The numerator of $h$ and its derivative in
$\rho_a$ vanish at $\rho_a=0$ for every $\bd$; the derivative is
$(w\bd+r_\e)/w-\bd-r_\e/w$. Hence $h$ is a smooth function of
$(\rho_a,\bd,w,r_\e)$ for $\rho_a\ge0$. At $\bd=0$,
$h=(\arctan\vartheta-\vartheta)/m^2$, which vanishes at $\rho_a=0$.

\emph{Yaw-rate offset.} Write $\arctan(b/a)-\delta_a=\vartheta+m^2h$ and
$\dot\rho_a=-k_1m+m^2\varpi$, with
$\varpi=-w\,\bd^2\,(1-\cos(m\bd))/(m\bd)^2+\bz E_1$. Then, by
\eqref{eq:rstar} and \eqref{eq:A-err},
\begin{equation}
G=-\partial_{\rho_a}\vartheta\;\varpi-\frac{2(1-\sigma)\dot\rho_a}{\rho_a}\,h-\dot h
+\Bigl(c-\frac{2(1-\sigma)\dot\rho_a}{\rho_a}\Bigr)\bz+\chi ,
\label{eq:A-G}
\end{equation}
in which $\dot\rho_a/\rho_a=(1-\sigma)(-k_1+m\,\varpi)$ and
$\dot h=\partial_{\rho_a}h\,\dot\rho_a+\partial_{\bd}h\,\dot{\bd}+\partial_th$
are smooth in $\Theta$ by \eqref{eq:t1}, \eqref{eq:t2}. Every term of
\eqref{eq:A-G} vanishes at $\rho_a=0$, $\bd=\bz=0$. Hence, with continuous
nondecreasing $L_1$, $L_2$ that depend on the leader only through the
bounds on $w$, $\dot w$, $r_\e$, $\dot r_\e$,
\begin{equation}
\begin{aligned}
|h|&\le L_1(\rho_a+|\bd|)\,(\rho_a+|\bd|),\\
|G|&\le L_2(\rho_a+|\bd|+|\bz|)\,(\rho_a+|\bd|+|\bz|).
\end{aligned}
\label{eq:A-L}
\end{equation}

\emph{Estimate of $M$.} By \eqref{eq:znr} and \eqref{eq:Md2},
$\bar r-\bar r^\ast=\tilde r+\kappa_\e\varpi$, so that
$M=\rho_a+|\bd|+|\tilde\theta|+|\tu|+|\tv|+|\tilde r+\kappa_\e\varpi|$, and
$|\varpi|\le L_3(|\Theta|)\,(|\bd|+|\bz|)$ with $L_3$ continuous and
nondecreasing. By \eqref{eq:A-hG} and \eqref{eq:A-L}, $|\Theta|+s$ and $M$
are each bounded by a class-$\mathcal K_\infty$ function of the other, and
\eqref{eq:A-k1}, \eqref{eq:A-s} give \eqref{eq:dclaim1},
\eqref{eq:kclaim1} and \eqref{eq:uclaim}.

\emph{Hull distance.} By \eqref{eq:qx}, \eqref{eq:qy}, the vector between
the pivot points differs from the vector between the zero-sideslip points
by a vector of length at most $D_\e|\theta|$ and by the vector with the
components $(X_1/Y_0)\tu$ and $-\tv/Y_0$. Hence
$\rho\le\rho_a+D_\e|\theta|+(|X_1|/|Y_0|)|\tu|+|\tv|/|Y_0|$. With
$|\theta|\le m^2|\tilde\theta|+|\vartheta|$, $m\le\ell$ and
$|\vartheta|\le\rho_a\,\omega_M/u_{\min}$, this gives \eqref{eq:kclaim2}
and \eqref{eq:dclaim2}.

\emph{Inputs.} The feedback divides by $\rho_a$, and its inputs are
bounded because they are smooth functions of the normalized errors. By
\eqref{eq:A-err} and \eqref{eq:A-hG},
$r_\p-r_\e=\Gamma+m^2(\bzeta+G)$, with
$\Gamma=r_{\rm ref}-r_\e=-(\partial_t-k_1m\,\partial_{\rho_a})\vartheta$.
Since $\vartheta$ and $m$ carry the factor $\rho_a$,
$|\Gamma|\le L_4\,\rho_a$ and $|\partial_{\rho_a}\Gamma|\le L_4$, where
$L_4$ depends on the bounds on $w$, $r_\e$ and their first derivatives.
Under the hypotheses of Theorem~\ref{thm:dyn}, also
$|\partial_t\Gamma|\le L_4'\,\rho_a$, where $L_4'$ depends in addition on
the bounds on the second derivatives of $w$ and $r_\e$. Hence
$|r_\p-r_\e|\le L_5(|\Theta|)\,|\Theta|$, which is \eqref{eq:kclaim3},
and, with $|\sqrt{a^2+b^2}-w|\le k_1m+|b|$ and
$|b|\le L_5(|\Theta|)\,|\Theta|$, \eqref{eq:uclaim2}. By \eqref{eq:tauu},
\eqref{eq:taur}, \eqref{eq:z4} and \eqref{eq:w},
\begin{subequations}\label{eq:A-tau}
\begin{align}
\tau_u-\tau_u^\e&=f_u(u_\e,v_\e,r_\e)-f_u(u_\p,v_\p,r_\p)\notag\\
&\quad+\frac{Y_0}{X_1}\,(u_a-w-\tu)-\frac{v_\p\,r_\p-v_\e\,r_\e}{X_1},\label{eq:A-tu}\\
\tau_r-\tau_r^\e&=f_r(u_\e,v_\e,r_\e)-f_r(u_\p,v_\p,r_\p)\notag\\
&\quad+\frac{\dd}{\dd t}\bigl(\Gamma+m^2G\bigr)\notag\\
&\quad-\Bigl(k_r-\frac{2(1-\sigma)\dot\rho_a}{\rho_a}\Bigr)m^2\bzeta+m^2\bz .\label{eq:A-tr}
\end{align}
\end{subequations}
Along the closed loop,
$\dot\Gamma=\partial_t\Gamma+\partial_{\rho_a}\Gamma\,\dot\rho_a$ with
$\dot\rho_a=-k_1m+m^2\varpi$, and
$\frac{\dd}{\dd t}(m^2G)=2m\,\dot m\,G+m^2\dot G$, in which $\dot m$ and
$\dot G$ are bounded by \eqref{eq:target} and by the bounds on the second
derivatives of $w$ and $r_\e$. Each term of \eqref{eq:A-tau} is therefore
bounded by $L_6(|\Theta|+s)\,(|\Theta|+s)$, which gives
\eqref{eq:dclaim3}. \hfill$\square$

\section{Proof of Proposition \ref{prop:par}}\label{app:par}

By \eqref{eq:A-err} and Theorems \ref{thm:kin}, \ref{thm:dyn},
$\theta=-\vartheta+m^2\tilde\theta$ with $\tilde\theta\to0$,
$m\le\rho_a$ and $\vartheta/\rho_a=r_\e/(w+k_1m)$; hence \eqref{eq:slopea}.

For the hull distance, the speed errors are divided by the range. The
quotients obey a linear system whose forcing converges, and this
determines the ratio of the hull distance to the distance between the
zero-sideslip points. Let $\hat U=\tu/\rho_a$ and $\hat V=\tv/\rho_a$. By \eqref{eq:t4},
\eqref{eq:t5},
\begin{subequations}\label{eq:B-UV}
\begin{align}
\dot{\hat U}&=-\Bigl(\frac{Y_0}{X_1}+\frac{\dot\rho_a}{\rho_a}\Bigr)\hat U-\frac{r_\p}{X_1}\,\hat V\notag\\
&\quad+\frac{Y_0}{X_1}\,\frac{u_a-w}{\rho_a}-\frac{v_\e}{X_1}\,\frac{r_\p-r_\e}{\rho_a},\label{eq:B-U}\\
\dot{\hat V}&=\Bigl(Y_0-\frac{\dot\rho_a}{\rho_a}\Bigr)\hat V+X_1\,r_\p\,\hat U+X(u_\e)\,\frac{r_\p-r_\e}{\rho_a}.\label{eq:B-V}
\end{align}
\end{subequations}
This system is linear in $(\hat U,\hat V)$, with coefficients that are
functions of $t$ along the solution $\Theta(t)$. As $t\to\infty$:
$\dot\rho_a/\rho_a\to-k_1$, by the expression after \eqref{eq:A-G};
$r_\p-r_\e\to0$; $(u_a-w)/\rho_a\to k_1$, because $a-w=k_1m$ and
$b=O(\rho_a)$; and $(r_\p-r_\e)/\rho_a-g\to0$, because
$r_\p-r_\e=(r_{\rm ref}-r_\e)+m^2(\bzeta+G)$ and
$(r_{\rm ref}-r_\e)/\rho_a=-\partial_t\vartheta/\rho_a+k_1(m/\rho_a)\,\partial_{\rho_a}\vartheta$.
For $Q=X_1^2\hat U^2+\hat V^2$ the terms with $\hat U\hat V$ cancel, and
the homogeneous part of \eqref{eq:B-UV} gives
$\tfrac12\dot Q=-X_1^2(Y_0/X_1+\dot\rho_a/\rho_a)\hat U^2+(Y_0-\dot\rho_a/\rho_a)\hat V^2$,
whose rates converge to $Y_0/X_1-k_1$ and $-Y_0-k_1$, both at least
$\mu-k_1>0$. The same holds for \eqref{eq:Ueq}, \eqref{eq:Veq}. Hence
$(\hat U,\hat V)$ is bounded and $(\hat U-U,\hat V-V)\to0$.

By \eqref{eq:qx}, \eqref{eq:qy},
\begin{subequations}\label{eq:B-rho}
\begin{align}
\frac{\rho\cos\gamma}{\rho_a}&=\cos\gamma_a+\frac{X(u_\e)}{Y_0}\,\frac{1-\cos\theta}{\rho_a}
+\frac{v_\e}{Y_0}\,\frac{\sin\theta}{\rho_a}+\frac{X_1}{Y_0}\,\hat U,\label{eq:B-c}\\
\frac{\rho\sin\gamma}{\rho_a}&=\sin\gamma_a+\frac{X(u_\e)}{Y_0}\,\frac{\sin\theta}{\rho_a}
-\frac{v_\e}{Y_0}\,\frac{1-\cos\theta}{\rho_a}-\frac{\hat V}{Y_0}.\label{eq:B-s}
\end{align}
\end{subequations}
With $\gamma_a\to0$, \eqref{eq:slopea} and
$1-v_\e r_\e/(Y_0w)=(Y_0u_\e+X_1\dot u_\e)/(Y_0w)$, the right side of
\eqref{eq:B-c} differs from $(Y_0u_\e+X_1\dot u_\e)/(Y_0w)+(X_1/Y_0)U$, and
that of \eqref{eq:B-s} from $-(X(u_\e)\kappa_\e+V)/Y_0$, by terms that
converge to zero. Hence $\rho/\rho_a-P\to0$, and \eqref{eq:slope} follows
from \eqref{eq:slopea}. \hfill$\square$

\section{Proof of Lemma \ref{lem:vl} and Theorem \ref{thm:vl}}\label{app:vl}

The virtual leader obeys the equations of the follower's hull.
Lemma~\ref{lem:vl} bounds its heading offset $\eta$ by an invariant strip,
and Theorem~\ref{thm:vl} then follows from Appendix~\ref{app:proof} with
the leader's signals replaced.

\emph{Proof of Lemma~\ref{lem:vl}.} By \eqref{eq:vl-sway},
$|v_\e(t)|\le b_\e\omega_M$ for all $t\ge0$, so, by \eqref{eq:off},
$|v_o|\le b_o\,\omega_M$ and
$|\dot v_o|=|Y_\e v_\e+B_\e r_\e+\dot r_\e\,d\cos\beta|\le2B_o\,\omega_M$.
If $\omega_M=0$, then
$v_o=\eta=0$, $w_d=u_o+(X_1/Y_0)\dot u_o$ and $r_d=0$. Let $\omega_M>0$
and consider the strip $|\eta|\le K_\eta\omega_M$. On it, by
\eqref{eq:vl-ud},
$|u_d-u_o|\le u_o\eta^2/2+|v_o||\eta|\le K_u\omega_M^2$. Since
$K_w\ge K_u\ge u_m\,K_\eta^2/2$ and $w_m\le u_m$,
\eqref{eq:vl-om1} gives $K_\eta\,\omega_M\le1$,
$K_u\,\omega_M^2\le K_w\,\omega_M^2\le w_m/2$ and
$|1+X_1|\,K_u\,\omega_M^2\le d_m/2$. Hence, on the strip,
$u_d+X(u_d)=u_o+X(u_o)+(1+X_1)(u_d-u_o)\ge d_m/2$ and
$|X(u_d)|\le X_M+|X_1|u_m/2$. The numerator of \eqref{eq:vl-gen}
is
$N=(Y_0u_o-\dot u_o)\sin\eta+(Y_\e-Y_0)\,v_\e\cos\eta
+\bigl[B_\e\cos\eta-X(u_d)\bigr]r_\e
{}+(\dot r_\e-Y_0\,r_\e)\,d\cos\beta\cos\eta$,
in which $Y_0u_o-\dot u_o\le-h_0$, and the last three terms are bounded
by $C_0\omega_M$. At $\eta=K_\eta\omega_M$, $\sin x\ge5x/6$ for
$0\le x\le1$ gives
$N\le-\frac56h_0K_\eta\omega_M+C_0\omega_M=-\frac23C_0\omega_M<0$, and at
$\eta=-K_\eta\omega_M$, $N\ge\frac23C_0\omega_M>0$. The strip is therefore
invariant, the right side of \eqref{eq:vl-gen} is bounded and locally
Lipschitz in $\eta$ on it, and $\eta$ exists for all $t\ge0$. On the
strip, $|N|\le\bigl((|Y_0|u_M+a_M)K_\eta+C_0\bigr)\omega_M$, so
$|\dot\eta|\le K_r\omega_M$ and $|r_d|\le(1+K_r)\omega_M$. Finally
$|v_d|\le(u_M\,K_\eta+b_o)\omega_M$, and \eqref{eq:vl-wd} is
$w_d=u_d+(X_1/Y_0)\bigl[\dot u_o\cos\eta+\dot v_o\sin\eta\bigr]
+(v_d/Y_0)\,[r_\e+(1+X_1)\dot\eta]$. Hence
$|w_d-u_o-(X_1/Y_0)\dot u_o|\le K_w\omega_M^2\le w_m/2$, and
$w_m\le u_o+(X_1/Y_0)\dot u_o\le u_M+|X_1|a_M/|Y_0|$.
\hfill$\square$

\emph{Proof of Theorem~\ref{thm:vl}.} Differentiating \eqref{eq:vl-ud},
\eqref{eq:vl-vd} with \eqref{eq:vl-gen} gives
$\dot v_d=Y_0v_d+X(u_d)\,r_d$ and
$\dot u_d=\dot u_o\cos\eta+\dot v_o\sin\eta+v_d\,\dot\eta$. The virtual leader
therefore satisfies the kinematics and the sway equation of the follower's
hull, with the surge speed $u_d(t)$ and the yaw rate $r_d(t)$. Its
zero-sideslip point, displaced from the offset point by
$X(u_d)/Y_0$ along the heading $\psi_\e+\eta$ and by $-v_d/Y_0$ across it,
moves along that heading at the speed
$u_d+v_dr_d/Y_0+(X_1/Y_0)\,\dot u_d$, which is \eqref{eq:vl-wd}. Hence
\eqref{eq:zmodel} holds for the polar variables of the zero-sideslip
points of the follower and of the virtual leader, with $(w,r_\e)$ replaced
by $(w_d,r_d)$. Since \eqref{eq:vl-wd} is
$\dot u_d=(Y_0/X_1)(w_d-u_d)-v_dr_d/X_1$, the speed errors $u_\p-u_d$,
$v_\p-v_d$ obey \eqref{eq:t4}, \eqref{eq:t5} with
$(u_\e,v_\e,r_\e,w)$ replaced by $(u_d,v_d,r_d,w_d)$.

The proof in Appendix~\ref{app:proof} uses the leader only through these
equations, through the bounds on $w$, $r_\e$, $v_\e$ and $X(u_\e)$, and
through the continuity of $w$ and $r_\e$ and the boundedness of their
derivatives. By Lemma~\ref{lem:vl}, on the invariant strip $\dot\eta$ is a
smooth bounded function of $(\eta,u_\e,\dot u_\e,v_\e,r_\e)$, and of
$\dot r_\e$ if $d>0$, with bounded
derivatives. Hence $r_d$ and $w_d$ are continuous, their derivatives are
bounded and piecewise continuous when $\ddot u_\e$ and $\dot r_\e$ are,
and $\ddot r_\e$ if $d>0$, and
\eqref{eq:vl-b} supplies the bounds, with $u_{\min}$, $u_{\max}$,
$\omega_M$ replaced by $w_m/2$, $u_M+|X_1|a_M/|Y_0|+w_m/2$,
$(1+K_r)\,\omega_M$. With surge force and yaw moment inputs, $\dot r_d$ is
continuous with a piecewise continuous derivative when $\ddot u_\e$ and
$\dot r_\e$ are, and $\ddot r_\e$ if $d>0$. The proof of Appendix~\ref{app:proof} therefore holds with
the replacement \eqref{eq:vl-rule}. The pivot point of the virtual leader
is the offset point, so $\rho$ is the distance from the
follower's hull to the offset point, and \eqref{eq:kclaim2}, \eqref{eq:dclaim2} hold with $\theta$
replaced by $\theta-\eta$ and $C_\rho$ by another constant.

The estimates of the inputs in Appendix~\ref{app:A4} use the leader
through the same equations and bounds, and, for \eqref{eq:dclaim3},
through the bounds on the second derivatives of $w$ and $r_\e$. On the
invariant strip, $r_d$ and $w_d$ are smooth functions of
$(\eta,u_\e,\dot u_\e,v_\e,r_\e)$, and of $\dot r_\e$ if $d>0$, so their second derivatives are bounded
when $\dddot u_\e$ and $\ddot r_\e$ are, and $\dddot r_\e$ if $d>0$. Since
$\dot u_d=(Y_0/X_1)(w_d-u_d)-v_dr_d/X_1$, the identities \eqref{eq:A-tau}
hold with $(\tau_u^\e,\tau_r^\e)$ replaced by $(\tau_u^d,\tau_r^d)$.
Hence \eqref{eq:kclaim3} and \eqref{eq:dclaim3} hold with the replacement
\eqref{eq:vl-rule}. \hfill$\square$

\vspace*{-13pt}\paragraph*{Acknowledgment.} The  problem and result building are the authors. Some mathematical derivations were carried out by Claude under the authors' strict direction, correction and 
verification.

\end{document}